\documentclass[a4paper,fleqn,usenatbib]{mnras}
\usepackage{graphicx}	
\usepackage{amsmath}	
\usepackage{amssymb}	

\def\apjs{{\rm ApJS}}                  
\def\apjl{{\rm ApJ Let}}               
\def\aj{{\rm AJ}}                      
\def\mnras{{\rm MNRAS}}                        

\def\Msol{\thinspace\hbox{$\hbox{M}_{\odot}$}}

\def\a4{\hsize 17.0cm \vsize 25.cm}
\newcommand{\der}[2]  { \frac{{\rm d}#1}{{\rm d}#2} }

\newcommand{\dif}     {{\rm d}}

\title[Early evolution of SSCs]{On the early evolution of massive star clusters: the case of  
         cloud D1 and its embedded cluster in NGC 5253}

\author[S. Silich et al.]
{Sergiy Silich \thanks{E-mail: silich@inaoep.mx}$^{1},$
    Guillermo Tenorio-Tagle$^{1},$
    Sergio Mart\'{\i}nez-Gonz\'alez$^{2},$ 
    Jean Turner$^{3}$
\\
$^{1}$Instituto Nacional de Astrof\'\i sica \'Optica y Electr\'onica, AP 51, 
      72000 Puebla, M\'exico\\
$^{2}$CONACYT-Instituto Nacional de Astrof\'isica, \'Optica y Electr\'onica, AP 51, 
      72000 Puebla, M\'exico\\
$^{3}$Department of Physics and Astronomy, UCLA, Los Angeles CA 90095-1547 USA\\
}

\date{Accepted XXX. Received YYY; in original form ZZZ}

\pubyear{2016}

\begin{document}
\label{firstpage}
\pagerange{\pageref{firstpage}--\pageref{lastpage}}
\maketitle

\begin{abstract}
We discuss a theoretical model for the early evolution of massive star clusters and confront it
with the ALMA, radio and infrared observations of the young stellar cluster highly obscured by 
the molecular cloud D1 in the nearby dwarf spheroidal galaxy NGC 5253. We show that a large
turbulent pressure in the central zones of D1 cluster may cause individual wind-blown bubbles to reach 
pressure confinement before encountering their neighbors.  In this case stellar winds are added to the 
hot shocked wind pockets of gas around individual massive stars that leads them to meet and produce a 
cluster wind in time-scales less than $10^5$ yrs. In order to inhibit the possibility of cloud dispersal, or 
the early negative star formation feedback, one should account for mass loading that may come, for 
example, from pre-main sequence (PMS) low-mass stars through photo-evaporation of their proto-stellar 
disks. Mass loading at a rate in excess of 8$\times 10^{-9}$ M$_{\odot}$ yr$^{-1}$ per each PMS 
star is required to extend the hidden star cluster phase in this particular cluster. In this regime, the 
parental cloud remains  relatively unperturbed, while pockets of molecular, photoionized and hot gas 
coexist within the star forming region. Nevertheless, the most likely scenario for cloud D1 and its 
embedded cluster is that the hot shocked winds around individual massive stars should merge at an 
age of a few millions of years when the PMS star proto-stellar disks vanish and mass loading ceases 
that allows a cluster to form a global wind.
\end{abstract}


\section{Introduction}
\label{sec1}

Most stars form in clusters. However only a small fraction of these clusters survives for a long 
time (Gyrs) as globular clusters (GCs) \citep[see][]{Zwart2010}. It was suggested that most massive 
young stellar clusters observed in our local group of galaxies may constitute present-day 
proto-globular clusters \citep{Elmegreen1997}.  However so far it is not clear if multiple stellar 
populations with peculiar chemical patterns  \citep[e.g.][]{Lee1999,Bedin2004,Marino2008,
Carretta2009,Renzini2015} are formed in present-day massive clusters as it occurred long ago in globular 
clusters or in this respect present-day and globular cluster populations are different and thus star 
formation  deeply depends on their environment conditions.

There are two major scenarios of how star formation proceeds in star cluster progenitor clouds 
\citep[e.g.][]{Longmore2014, Banerjee2015}. In an in situ, monolithic formation scenario stellar 
clusters form in extremely dense and compact progenitor clouds. In a hierarchical or conveyor-belt 
scenario star formation occurs in the densest peaks spread over the natal pre-stellar cloud. The 
centrally concentrated cluster is then formed due to infalling and merging of such subclusters in the 
central zone of the star forming region. 

During their early evolution star clusters remain deeply embedded into their parental molecular 
clouds and are not observable in the visible line regime \citep[see a phenomenological 
classification scheme by][]{Whitmore2011}. However, stellar winds and the ionizing radiation from 
massive stars are believed to play a major role in star cluster formation \citep[e.g.][]{Krause2012,
Dale2015,Calura2015,Rahner2017}. Indeed, neighboring wind 
collisions and the residual gas ionization by massive stars boost the gas pressure and result in the 
formation of a global star cluster wind, the residual gas expulsion from the star forming region 
(the process, which is usually referred as a negative star formation feedback) and the young cluster 
emerging in the UV and visible line regimes. 

The impact of negative stellar feedback on the parental molecular cloud is crucial as the 
residual gas contributes significantly to the gravitational field of star forming clouds and 
its rapid expulsion promotes cluster dissolution in a short time-scale \citep[infant 
mortality, see][]{Boily2003,Baumgardt2007,Zwart2010}  that does not allow the long-lived, 
proto-globular clusters to form. 

Some observations strongly support this scenario. For example, \citet{Bastian2014,Bastian2015B} 
were unable to find any significant amount of residual gas in their sample of  young massive 
clusters and concluded that even high-mass clusters expel the natal gas within a few Myrs after 
their formation. High resolution survey of the molecular gas distribution around star forming
regions in Large Magellanic Cloud led \citet{Seale2012} to conclude that young clusters
destroy their natal molecular clouds on a time-scale of at least a few times $10^5$yr.
This conclusion strongly restricts scenarios for globular and bound clusters formation. 

However, recent observations of nearby galaxies in infrared, millimeter and radio 
wavelengths revealed an obscured mode of star formation in such galaxies as SBS 0335-052 
\citep{Thuan1999}, He 2-10 \citep{Vacca2002}, NGC 2366 \citep{Oey2017} and NGC 5253 
\citep{Turner2015,Beck2015, Turner2017,Cohen2018}. The extension of the hidden star cluster
phase is also required in some models of the multiple stellar generations formation in globular 
clusters \citep[e.g.][]{Kim2018,TenorioTagle2019}.

Thus, the current understanding of the physical processes involved in different star cluster formation 
scenarios remains incomplete. The disruption of natal gas clouds by stellar winds and the residual 
gas expulsion at early stages of the star cluster assembling, the 
characteristic time-scale for the global star cluster wind development and mechanisms which may 
suppress or delay the progenitor cloud disruption are still between the intensively debated 
problems \citep[see][]{Krause2012,Calura2015,TenorioTagle2015,Krause2016,
Silich2017,Silich2018,Wunsch2017,Szecsi2019}.

Recent hydrodynamical simulations of gas-rich galaxies merging allowed one to model star
cluster formation down to 4\Msol \, and 0.1pc space resolution \citep{Lahen2019A,Lahen2019B}. 
However, the disruptive effects of massive star winds on the star forming cloud and the impact 
of low mass star outflows on the gas dynamics are still beyond the scope of these simulations.

\citet{Silich2017,Silich2018} discussed the interplay among young massive stars and the residual 
gas in young stellar clusters semi-analytically that allowed them to study stellar winds dynamics 
at much smaller scales. They concluded that in compact and massive star-forming clouds 
wind-driven shells around individual massive stars may stall before merging and hot shocked winds 
may be pressure confined due to a high intra-cloud gas pressure and strong radiative cooling. This 
suppresses star cluster winds and prevents the residual gas expulsion from the star-forming cloud. 
However, the model predictions are sensitive to the star cluster size, which is the 
``hardest parameter to measure in extragalactic sources'' \citep{Beck2015}.  Therefore, one has to 
confront model predictions with observations of young stellar clusters in nearby galaxies provided 
with an extreme spatial resolution.

Probably, the best example of such a cluster is a  young stellar cluster in the nearby dwarf 
spheroidal galaxy NGC 5253 which is deeply embedded into a dense molecular cloud D1 and 
coincides with an extremely bright in radio and IR wavebands supernebula 
\citep{Turner2000,Turner2004,Gorjian2001,AlonsoHerrero2004,Beck2015}. The NGC 5253 
radio continuum is almost entirely thermal \citep{Beck1996} that indicates the youth of NGC 
5253 clusters, unless their most massive stars collapsed directly into black holes without 
exploding as supernovae \citep[e.g.][]{Adams2017A,Adams2017B, Mirabel2017A,Mirabel2017B}.

The supernebula is very compact, with $\sim 0^".1$ FWHM that corresponds to 
$\approx 1.9$~pc at the distance of 3.8 Mpc adopted to NGC 5253  and extremely 
bright, accounting for about 1/2 of the entire galaxy IR luminosity \citep{Gorjian2001}. 
It is located in the center of a compact ($\sim 0^".3$ FWHM) molecular cloud (D1) with 
a virial mass within $R_{D1} = 2.8$~pc of $\approx 2.5 \times 10^5$\Msol 
\, \citep{Turner2015,Turner2017}. 

\citet[][]{Turner2000,Gorjian2001,Beck2015} argued that the supernebula is excited by a young 
massive cluster still enshrouded by the parental molecular cloud D. The age of the cluster inferred
by different authors falls in the range of 1Myr - 3.5Myr \citep[see][]{AlonsoHerrero2004,
MonrealIbero2010,Calzetti2015,Smith2016}.
This system offers an excellent opportunity to study the early stages of star cluster formation. 
\citet{Silich2017} compared this cluster and star-forming cloud with their model
and found that individual wind-driven shells cannot merge in this case. However, at that time high
resolution ALMA observation of NGC 5253 were still not available. New observation with about
10 times better spatial resolution \citep{Turner2017} revealed that cloud D, which was previously
identified as a star-forming cloud is actually composed of several molecular clouds and that cloud 
D1 associated with the radio/infrared supernebula is much less massive and more compact than 
cloud D previously considered as a parental cloud for the enshrouded cluster. The molecular gas 
mass in D1 is very uncertain due to the high CO excitation, but is unlikely to exceed 35\% of the cloud 
virial mass \citep{Turner2017}. Hereafter we adopt that in the central zone of cloud D1 with radius 
2.8~pc gas and stars contribute 35\% and 65\% to the  virial mass: $M_g(R_{D1})=8.75 \times 
10^4$\Msol \, and  $M_{\star}(R_{D1})=1.625 \times 10^5$\Msol , respectively.

The comparison of the \citet{Silich2018} model predictions with new observations led to conclude 
that wind-driven shells stall, but parcels of hot shocked winds around individual stars should
merge to form a global wind in a short time-scale \citep[less than 
$10^5$ years, see equations A8 and A9 in][]{Silich2018} while ALMA observations clearly 
indicate that molecular gas in cloud D1 is relatively undisturbed. This led us to reexamine several 
physical processes not considered in the previous calculations and notice that mass loading from 
pre-main sequence (PMS) stars may prevent the pockets of hot shocked gas around individual 
massive stars from merging and thus suppress/delay the global star cluster wind development and 
the residual gas expulsion from young star-forming systems. 

The paper is organized as follows. The model adopted for the residual molecular gas and stellar 
density distribution, the expected number of low mass PMS stars and their mass 
loss rates are discussed in Section 2. The gas dynamics in the shocked wind zones is discussed in 
Section 3. Here relevant hydrodynamic equations are formulated, different hydrodynamic 
regimes and the role of PMS stars are discussed. In this section we also discuss the hot gas volume 
filling factor, the ionized gas mass in the central zone of cloud D1 and model uncertainties and
simplifications. The expected cloud D1 evolution is discussed in Section 4. Our major results and 
conclusions are summarized in Section 5. 

\section{Model setup}
\label{sec2}

\subsection{The star-forming molecular cloud D1}

Molecular cloud D1 and the associated radio and IR supernebula parameters are derived from
Gaussian fits to the CO, radio and IR integrated intensity maps and expressed
in terms of FWHM \citep[see][]{Turner2000,Gorjian2001,Turner2015,Turner2017}. The molecular
gas distribution in a ``Firecracker'' molecular cloud in Antennae galaxies, a potential site to form a 
massive star cluster, can also be fit by a Gaussian profile  \citep{Finn2019}. Therefore  hereafter a 
Gaussian distribution for the residual gas and recently formed stars, which likely excite the radio and 
IR supernebula, is adopted. As the FWHMs for CO and radio/IR sources are different, we assume 
that the molecular gas has a shallower radial density profile than the stellar mass distribution
(core radii $a = FWHM_{CO} / 2.355 \approx 2.4$~pc and $b = FWHM_{Radio/IR} / 2.355 
\approx 0.8$~pc for the molecular gas and stars, respectively). This agrees with the comparison of
\citet{Walker2016} between the distribution of stars in young stellar clusters and the distribution 
of gas in the proto-cluster molecular clouds. The gas, the stellar mass distribution and the number of 
stars per unit volume as a function of distance from the molecular cloud center then are: 
\begin{eqnarray}
      \label{eq1a}
      & & \hspace{-0.9cm} 
\rho_g(r) = \frac{(1 - \epsilon) M_{tot}}{(2\pi)^{3/2} a^3} \exp{\left[-\frac{1}{2}
                   \left(\frac{r}{a}\right)^2\right]} ,
      \\[0.2cm] \label{eq1b}
      & & \hspace{-0.9cm} 
\rho_{\star}(r) =  \frac{\epsilon M_{tot}}{(2\pi)^{3/2} b^3} \exp{\left[-\frac{1}{2}
                   \left(\frac{r}{b}\right)^2\right]} ,
      \\[0.2cm] \label{eq1c}
      & & \hspace{-0.9cm} 
n_{\star}(r) =  \frac{N_{star}}{(2\pi)^{3/2} b^3} \exp{\left[-\frac{1}{2}
                   \left(\frac{r}{b}\right)^2\right]} ,                
\end{eqnarray}
where $M_{tot}$ is the total (stars and gas) mass of the star-forming cloud, $N_{star}$ is the total 
number of stars and $\epsilon$ is the global star formation efficiency. 

The gas ($M_g(r)$) and stellar ($M_{\star}(r)$) mass and the number of stars ($N_{star}(r)$) 
enclosed within a sphere of  radius $r$ are: 
\begin{eqnarray}
       \nonumber
      & & \hspace{-0.9cm} 
M_g(r) = (1-\epsilon) M_{tot} \left[erf \left(\frac{r}{2^{1/2} a}\right) - \right.
     \\[0.2cm]  \label{eq2a}
      & & \hspace{0.5cm}
         \left.      \left(\frac{2}{\pi}\right)^{1/2} \frac{r}{a} \exp\left[{-\frac{1}{2}
                   \left(\frac{r}{a}\right)^2}\right]\right] ,
     \\[0.2cm]  \nonumber
      & & \hspace{-0.9cm}
M_{\star}(r) = \epsilon M_{tot} \left[erf \left(\frac{r}{2^{1/2} b}\right) - \right.
    \\[0.2cm]  \label{eq2b}
      & & \hspace{0.5cm}
             \left.  \left(\frac{2}{\pi}\right)^{1/2} \frac{r}{b} \exp\left[-{\frac{1}{2}
                   \left(\frac{r}{b}\right)^2}\right]\right] ,
       \\[0.2cm] \nonumber
      & & \hspace{-0.9cm} 
N_{star}(r) = N_{tot} \left[erf \left(\frac{r}{2^{1/2} b}\right) - \right.
    \\[0.2cm]  \label{eq2c}
      & & \hspace{0.5cm}
              \left.  \left(\frac{2}{\pi}\right)^{1/2} \frac{r}{b} \exp\left[-{\frac{1}{2}
                   \left(\frac{r}{b}\right)^2}\right]\right] ,
\end{eqnarray}
where $erf(r)$ is the error function.  

The total mass of the cloud and the global star formation efficiency are determined by means of 
the masses $M_g(R_{D1})$ and $M_{\star}(R_{D1})$, radius $R_{D1}$ and the gas and stellar density 
distribution core radii $a$ and $b$:
\begin{eqnarray}
       \nonumber
      & & \hspace{-1.1cm} 
M_{tot} = \frac{M_g(R_{D1})}{erf \left(\frac{R_{D1}}{2^{1/2} a}\right) -
                 \left(\frac{2}{\pi}\right)^{1/2} \frac{R_{D1}}{a} \exp\left[{-\frac{1}{2}
                 \left(\frac{R_{D1}}{a}\right)^2}\right]}  + 
      \\[0.2cm]  \label{eq0a}
      & & \hspace{-0.2cm}
                 \frac{M_{\star}(R_{D1})}{erf \left(\frac{R_{D1}}{2^{1/2} b}\right) - 
                 \left(\frac{2}{\pi}\right)^{1/2} \frac{R_{D1}}{b} \exp\left[{-\frac{1}{2}
                 \left(\frac{R_{D1}}{b}\right)^2}\right]}
    \\[0.2cm]    \nonumber
      & & \hspace{-0.7cm}
\epsilon = \frac{M_{\star}(R_(D1)}{M_{tot}} \left[erf \left(\frac{R_{D1}}{2^{1/2} b}\right) - \right.
     \\[0.2cm]  \label{eq0b}
      & & \hspace{0.5cm}
       \left. \left(\frac{2}{\pi}\right)^{1/2} \frac{R_{D1}}{b} \exp\left[{-\frac{1}{2}
                   \left(\frac{R_{D1}}{b}\right)^2}\right]\right]
\end{eqnarray}
Equations (\ref{eq0a}) and (\ref{eq0b}) together with $R_{D1}$, $M_g(R_{D1})$ and 
$M_{\star}(R_{D1})$ obtained by \citet{Turner2017} give $M_{tot} = 4.64 \times 10^5$\Msol \,
and $\epsilon = 0.35$. 

The pressure profile in such a cloud is determined by the equation of the hydrostatic equilibrium 
\citep{Calura2015}:
\begin{equation}
      \label{eq3}
\der{P_g}{r} = - \frac{G M(r) \rho_g(r)}{r^2} , 
\end{equation}
where $G$ is the gravitational constant and $M(r) = M_g(r) + M_{\star}(r)$ is the total mass enclosed 
within a sphere of  radius $r$. Numerical integration of equation (\ref{eq3}) from a large radius where 
the gas pressure is negligible towards the center determines the gas pressure  $P_g(r)$ at different 
distances from the molecular cloud center. As in molecular clouds the gas pressure is dominated by 
turbulence \citep[e.g.][]{Elmegreen1997,Elmegreen2017,Johnson2015}, equation (\ref{eq3}) also allows
one to obtain the gas one-dimensional velocity dispersion $\sigma^2(r) = P_g(r)/\rho_g(r)$ 
\citep{Smith2001}. The distribution of the gas density, pressure and velocity dispersion in a Gaussian 
star-forming cloud with a total mass $M_{tot} = 4.64 \times 10^5$\Msol \, global star formation 
efficiency $\epsilon = 0.35$  and the gas and stellar mass distribution core radii $a = 2.4$pc and 
$b=0.8$pc,  is shown in Fig. 1.
\begin{figure}
\includegraphics[width=\columnwidth]{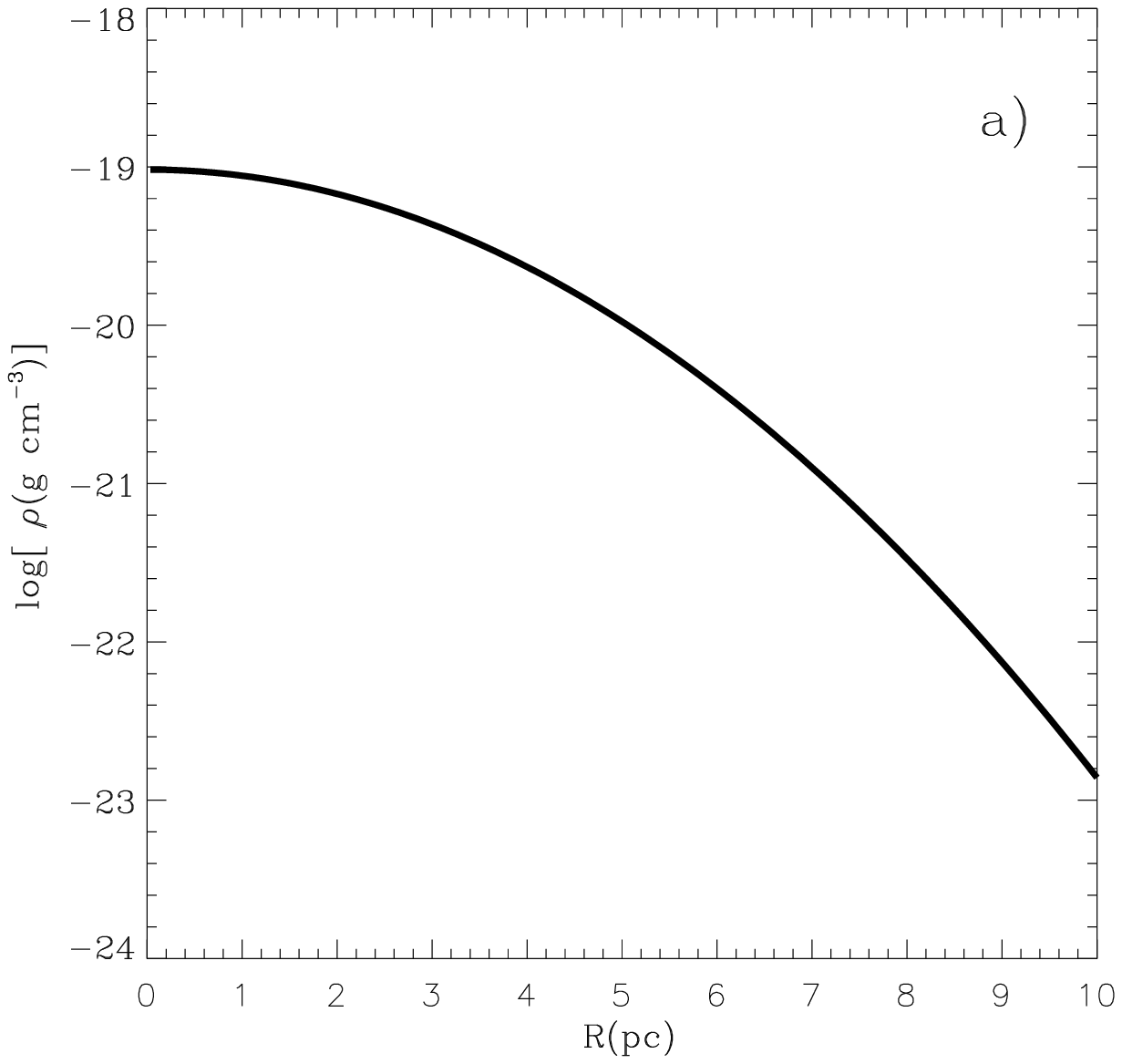}
\includegraphics[width=\columnwidth]{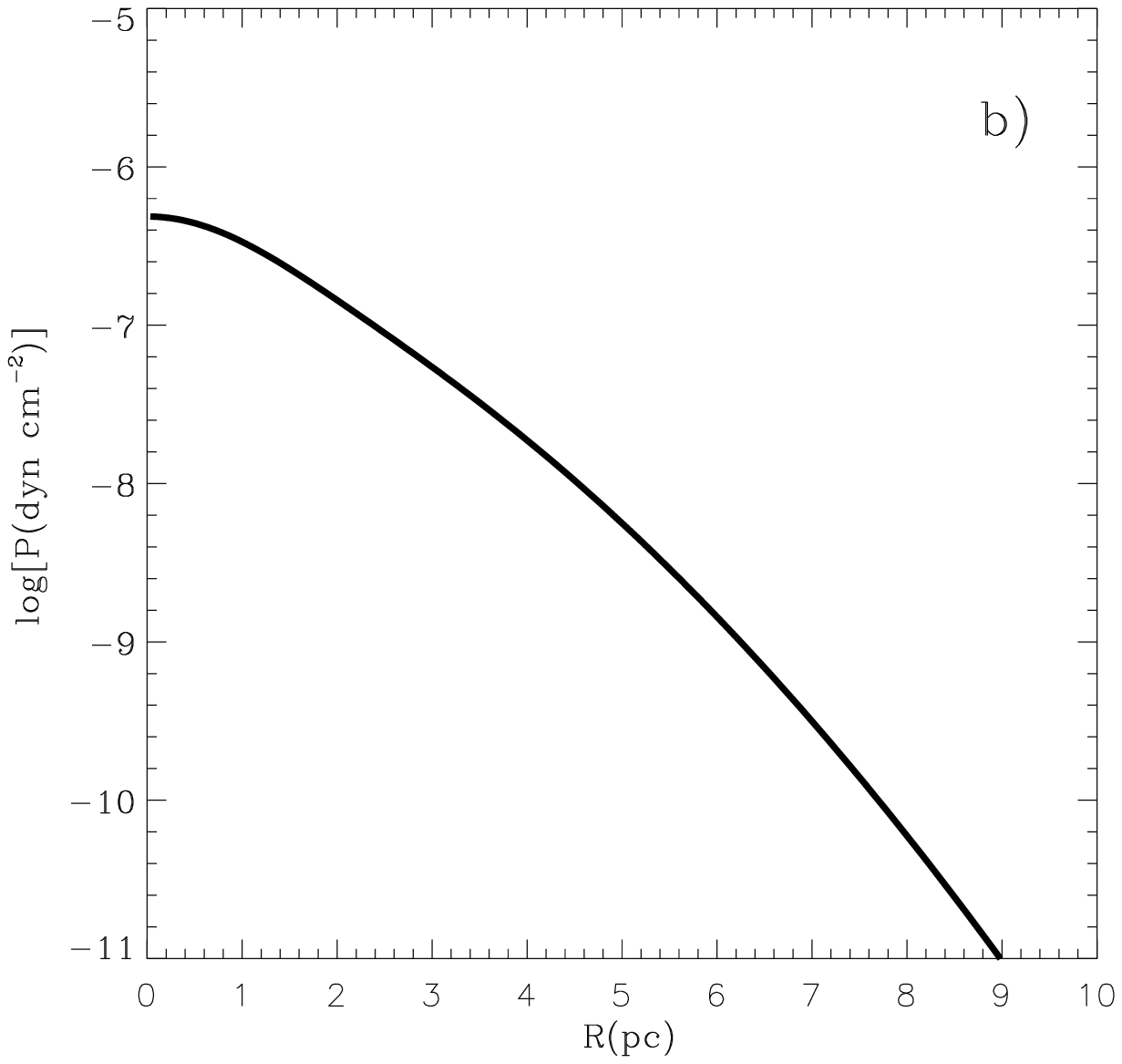}
\includegraphics[width=\columnwidth]{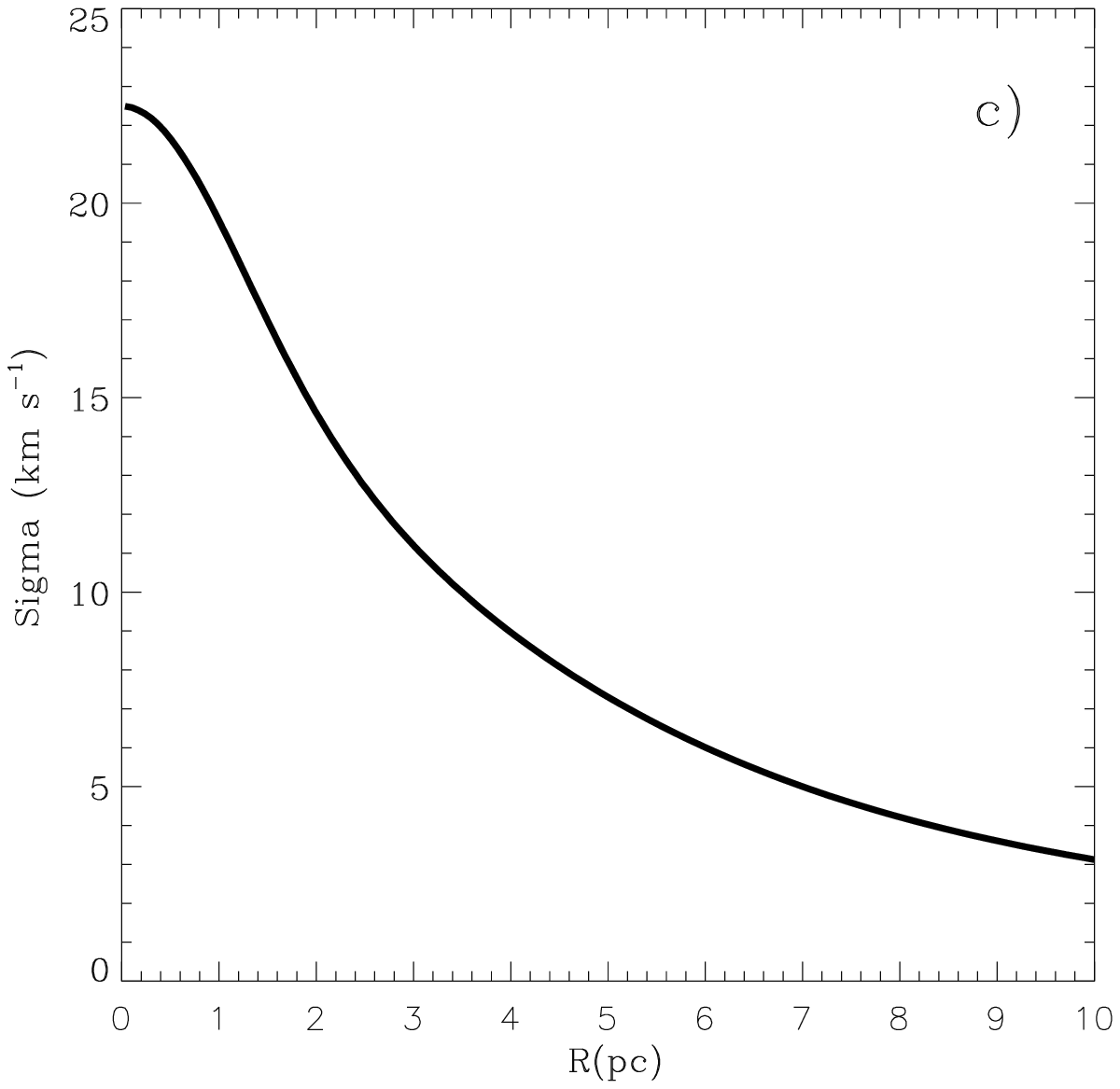}
\caption{The star-forming cloud. Panels a, b and c display the distributions of the gas density, pressure 
and 1D velocity dispersion, respectively.}
\label{f1}
\end{figure}

Following Starburst99 synthetic model we assume that in a young stellar cluster with a standard 
Kroupa initial mass function (IMF) the number of single massive ($M > 8$\Msol) stars $N_{mass}$ 
scales with the star cluster mass as 
\begin{equation}
      \label{eq4}
N_{mass} \approx 10^4 (M_{\star}/10^6\Msol) .
\end{equation}

The half distance $X$ between neighboring massive stars at different distances from the
star cluster center then is:
\begin{equation}
      \label{eq5}
X(r) = b \left[3 (\pi/2)^{1/2} \exp{\left[\frac{1}{2}\left(\frac{r}{b}\right)^2\right]}\right]^{1/3}
       N_{mass}^{-1/3} .
\end{equation}

\subsection{Massive stars mechanical luminosities and mass
             loss rates}

Hereafter it is assumed that all massive stars are identical. The representative star mechanical power 
$L_0$, mass loss and Lyman continuum rates ${\dot M}_0$ and $Q_0$ are then obtained by means
of Starburst99 version 7 synthetic model \citep{Leitherer1999}. Fig. 2 presents the time evolution of 
these quantities in the case of a standard Kroupa IMF, $Z = 0.2Z_{\odot}$ and the Padova stellar 
evolutionary tracks that include AGB stars. The representative stellar wind terminal
speed falls in the range of 2600km s$^{-1}$  -  2900km s$^{-1}$.
\begin{figure}
\includegraphics[width=\columnwidth]{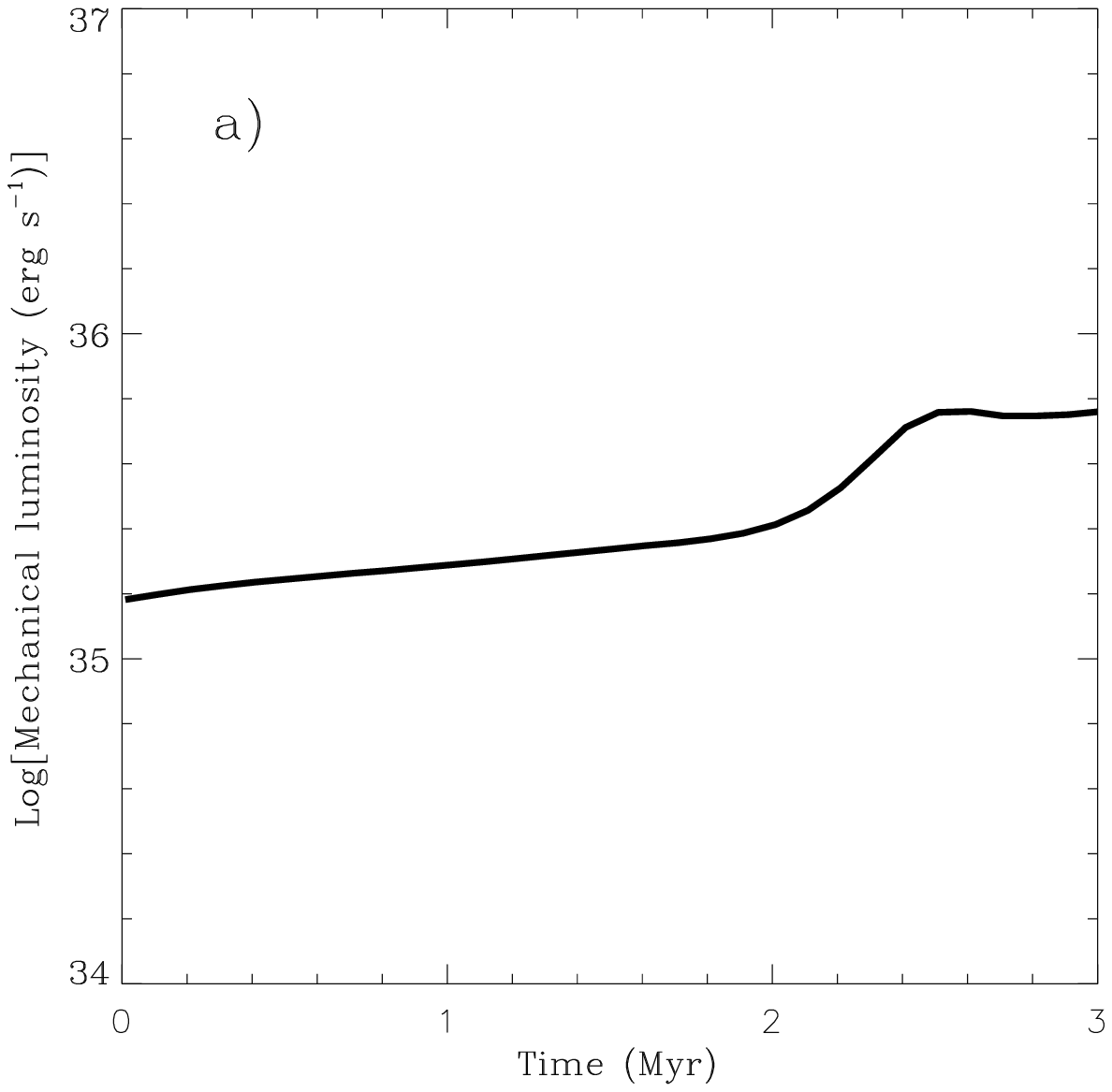}
\includegraphics[width=\columnwidth]{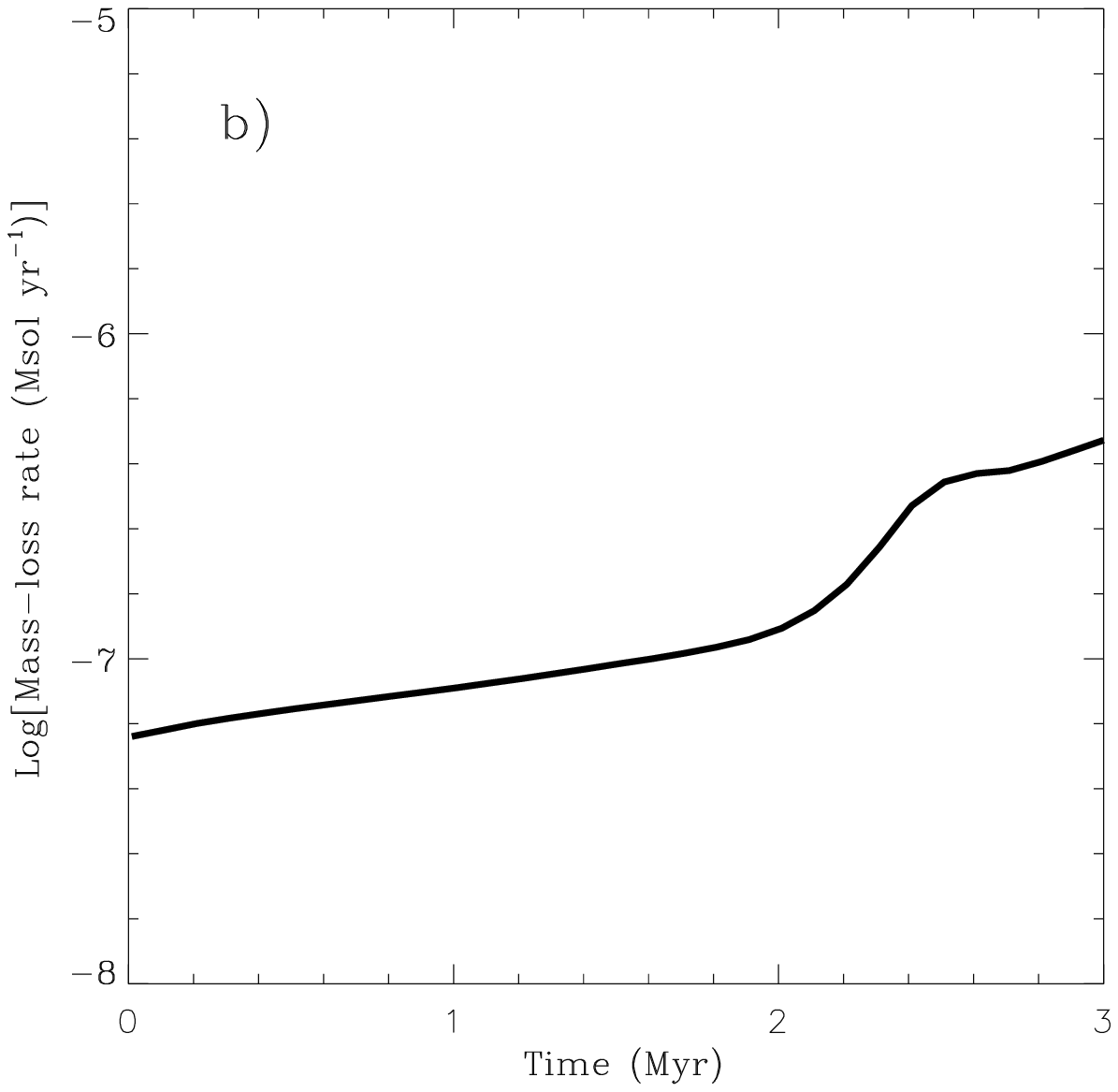}
\includegraphics[width=\columnwidth]{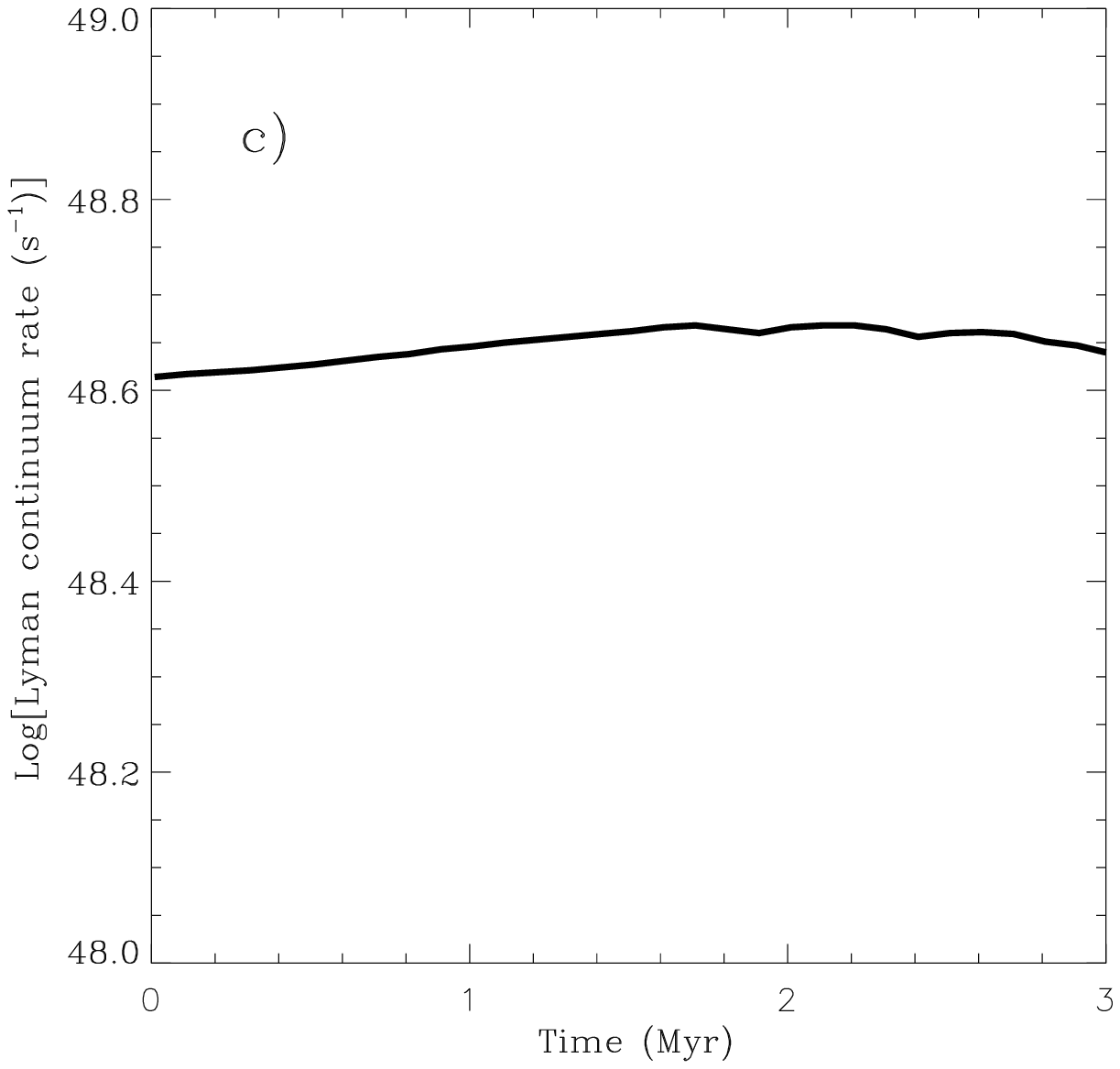}
\caption{Massive stars input rates. The mechanical power, mass loss and Lyman continuum rates of 
a representative massive star.}
\label{f2}
\end{figure}

\subsection{Pre-main sequence stars and their mass loss
            rates}

The number of PMS low mass ($M \le 3$~\Msol)  stars can be also scaled with the star cluster mass 
(see Appendix A):
\begin{equation}
      \label{eq6}
N_{PMS} \approx 1.5 \times 10^6 (M_{\star}/10^6\Msol) .
\end{equation}
This implies that in young stellar clusters more than 100 low mass PMS stars are located in  the 
$r \le X$ region around each massive star. These stars have proto-stellar disks which
lose mass continuously due to different processes, such as slow winds driven by the inward accretion
flows \citep{Appenzeller1989,Hartmann2018} or proto-stellar disk heating by the central low mass 
star X-ray emission \citep{Ercolano2008,Drake2009,Owen2011}, magnetically collimated, centrifugally 
launched jets, Herbig-Haro objects \citep[see][and references therein]{Bally2007,Coffey2008,
Frank2014,Nisini2018}.
However, it is likely that in young, compact clusters, where distances between massive and low 
mass stars are small, photoevaporation of proto-stellar disks by the neighboring massive star 
UV radiation dominates over the other mass-loss mechanisms \citep{Johnstone1998,Storzer1999,
Richling2000}. The proto-stellar disk mass-loss rates depend on many parameters, such as proto-stellar disk
orientation, mass, distance to the massive star, UV radiation intensity and in different observations and 
models fall into a wide range from $10^{-9}$\Msol \, yr$^{-1}$ to $10^{-6}$\Msol \, yr$^{-1}$. 
Analysis of pre-main sequence stellar populations in a number of stellar clusters shows that the 
proto-stellar disk fraction in the youngest star clusters exceeds 80\%, but it decreases with the star 
cluster age in a relatively short time-scale of  $\sim 3$~Myr \citep{Haisch2001,Lada2003,Fedele2010}.

\section{Subsonic shocked wind zones around individual 
              massive stars }
\label{sec3}

In massive and compact young stellar clusters wind-driven shells around individual stars stall before 
merging with their neighbors \citep{Silich2017} and disperse into a high pressure, turbulent intra-cloud 
medium. However, the hot shocked wind zones around massive stars continue to grow as their central 
stars supply mass and energy for a much longer time. Such hot blobs filled with shocked stellar wind gas 
grow in the subsonic regime until they merge with their neighbors or reach their cooling radii  
\citep{Silich2018}. 
\begin{figure}
\includegraphics[width=\columnwidth]{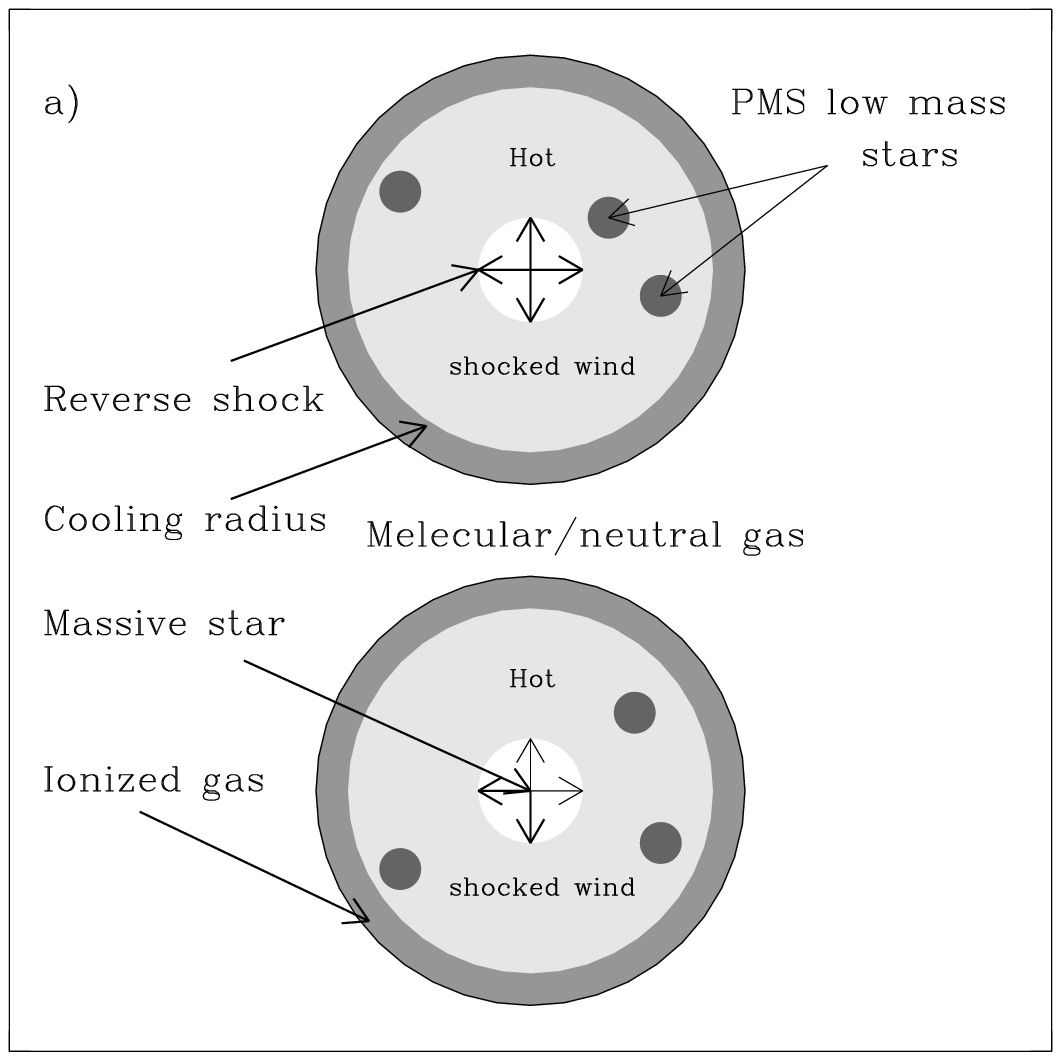}
\includegraphics[width=\columnwidth]{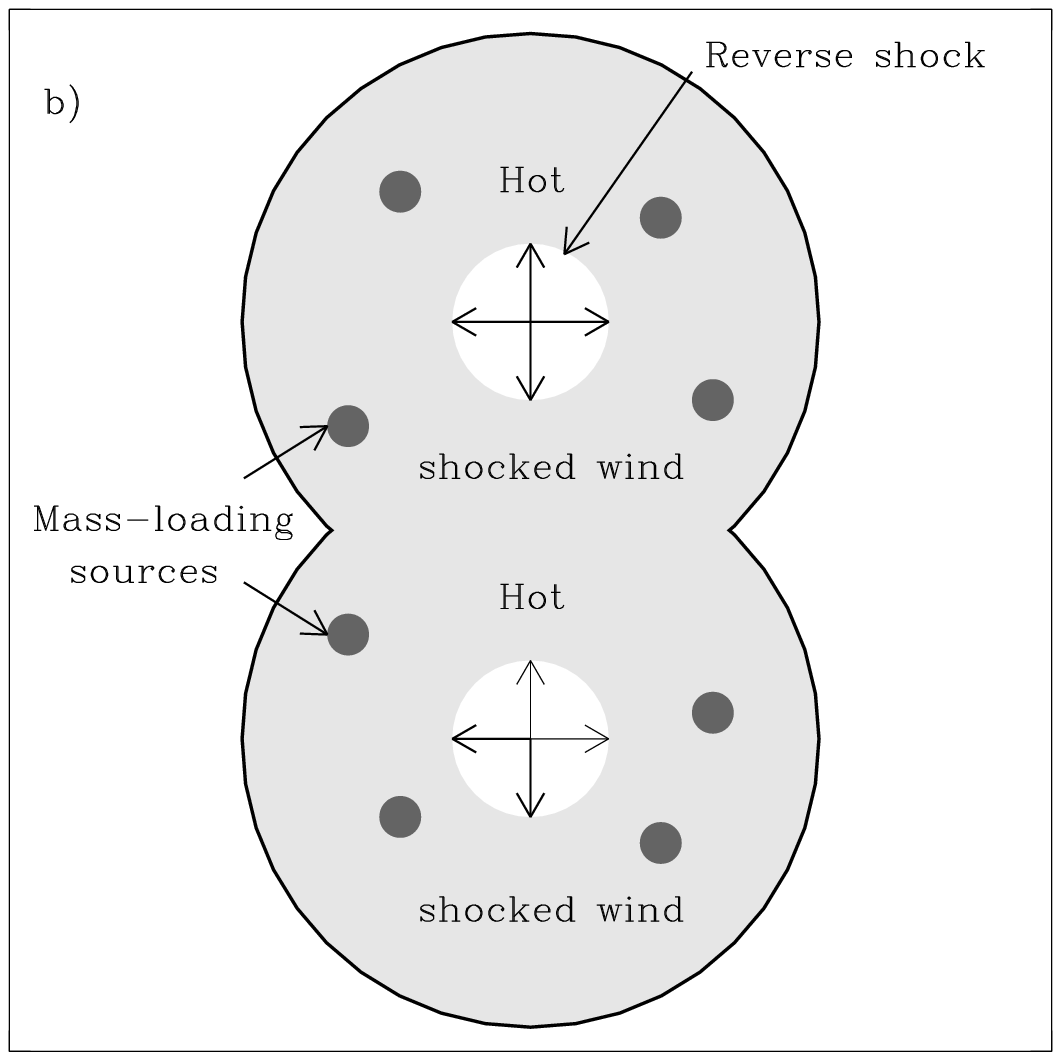}
\caption{Schematic representation of the gas distribution around individual massive stars. Panel a shows 
the gas distribution in the strong radiative regime, whereas panel b shows the same in the case when 
radiative cooling is unable to prevent the merging of hot neighboring shocked winds.}
\label{f3}
\end{figure}
In such shocked wind zones the reverse shock radius $R_{RS}$ is determined by the equation:
\begin{equation}
      \label{eq7}
R_{RS} = \left(\frac{L_0}{2 \pi V_{\infty} P_{ram}}\right)^{1/2} ,
\end{equation}
where $V_{\infty} = (2 L_0 / {\dot M}_0)^{1/2}$ is the stellar wind terminal speed (see
section 2.2) and the stellar wind ram pressure $P_{ram}$ is determined by the condition that 
the gas pressure at the edge of the shocked wind zone is equal to the turbulent pressure in 
the ambient medium.
In massive and compact star-forming molecular clouds the reverse shock radii $R_{RS}$ are small 
due to a large intra-cluster gas pressure. 
This leads to a large shocked gas density.  In addition, at early stages of evolution shocked winds 
could be mass-loaded via photoevaporation of PMS star proto-stellar disks. This would enhance the 
shocked gas density and boosts cooling rates in the shocked wind 
zones.  Thus a large intra-cluster gas pressure and mass loading may result in catastrophic shocked gas 
cooling at distances which are smaller than the half-distance between neighboring massive stars. In this 
regime shocked winds do not merge and pockets of hot gas remain immersed into cold, dense residual
gas. Under such conditions, star clusters do not form a global wind and do not expel the leftover gas into
the ambient ISM. Fig. 3 illustrates the two regimes.
In the catastrophic cooling regime (panel a) shocked mass-loaded winds around individual massive stars 
cool before merging with their neighbors, whereas in the lower pressure and less mass-loaded 
environments the neighboring shocked winds merge (panel b in Fig. 3), fill in the cluster volume with a 
hot gas and expel the residual gas into the ambient ISM. In the catastrophic cooling regime the hot, the
ionized and the molecular gas components live together, but as shown below, the ionized gas occupies 
only a small fraction of the star cluster volume whereas the size of the hot shocked wind zones
around individual massive stars and thus the hot component volume filling factor grows as 
the star cluster ages. 

\subsection{Main equations}

The distribution of the gas temperature, density and pressure in the subsonic shocked wind zones 
around individual massive stars is determined by the steady state, spherically symmetric 
hydrodynamic equations which include mass loading by PMS stars as a source term. As the 
number of low mass PMS stars around each massive star is large, we assume that mass loading 
is smoothly distributed inside shocked wind zones and normalize it to the number of PMS. 
The mass loading rate per unit space volume $q_m$ then is:
\begin{equation}
      \label{eq9}
q_m = {\dot M}_{PMS} \, n_{PMS} ,
\end{equation}
where ${\dot M}_{PMS}$ is the average mass loss rate per each PMS disk and $n_{PMS}$ is
the number of PMS stars per unit volume determined by equation (\ref{eq1c}) with $N_{star} = 
N_{PMS}$. ${\dot M}_{PMS}$ is a free parameter of the model and is assumed to be constant.
The set of the hydrodynamic equations which accounts for the mass loading and the shocked gas cooling 
then is \citep[e.g.][]{Silich2004}:
\begin{eqnarray}
      \label{eq9a}
      & & \hspace{-1.1cm} 
\frac{1}{r^2} \der{\rho u r^2}{r} = q_m , 
      \\[0.2cm]   \label{eq9b}  
      & & \hspace{-1.1cm}
\rho u \der{u}{r} = - \der{P}{r} - q_m u,
      \\[0.2cm]     \label{eq9c}
      & & \hspace{-1.1cm}
\frac{1}{r^2} \der{}{r}\left[\rho u r^2 \left(\frac{u^2}{2} +
\frac{\gamma}{\gamma-1} \frac{P}{\rho}\right)\right] = -Q ,
\end{eqnarray}
where $u, \rho$ and $P$ are the gas velocity, density and thermal pressure in the shocked gas
region, $r$ is the distance from the central massive star, $q_m$ is the mass loading rate.
$Q = n_i n_e \Lambda(T,Z)$ is the cooling rate, $\Lambda(T,Z)$ is the \citet{Raymond1976} 
cooling function, $n_i \approx n_e$ are the ion and electron number densities in the hot shocked 
wind blobs and $Z$ is the stellar wind metallicity. One can integrate equation (\ref{eq9a}) and present 
the set of  equations (\ref{eq9a})-(\ref{eq9c}) in a form suitable for the numerical integration:
\begin{eqnarray}
     \label{eq10a}
      & & \hspace{-1.1cm} 
\der{u}{r} = \frac{1}{r \rho} \frac{(\gamma-1) r Q + 2 \gamma u P - 
                    (\gamma+1) r q_m u^2 / 2}{u^2 - c^2} ,
      \\[0.2cm]     \label{eq10b}
      & & \hspace{-1.1cm}
\der{P}{r} = -\rho u \der{u}{r}  - q_m u,
      \\[0.2cm]     \label{eq10c}
      & & \hspace{-1.1cm}
\rho = \frac{{\dot M}_0}{4 \pi u r^2} + \frac{q_m R_{RS}^3}{3 u r^2} 
           \left(\frac{r^3}{R_{RS}^3}-1\right)  ,
\end{eqnarray}
where ${\dot M}_0$ is the central massive star mass loss rate, $R_{RS}$ is the reverse shock 
radius and $c^2 = \gamma P / \rho$ is the sound speed in the shocked wind plasma. The temperature of 
the shocked wind is $T = \mu_i c^2 / \gamma k$, where $\mu_i = 14/23 m_H$ is the mean mass per 
particle in the completely ionized gas with 1 helium atom per each 10 hydrogen atoms, $m_H$ is the 
mass of the hydrogen atom and $k$ is the Boltzmann constant. The set of equations (\ref{eq10a}) - 
(\ref{eq10c}) is integrated numerically outwards from the reverse shock position. The initial conditions 
for the numerical integration (the values of the shocked wind density, $\rho_{sw}$, temperature, 
$T_{sw}$, and pressure, $P_{sw}$ behind the reverse shock) are determined by the Rankine-Hugoniot 
conditions at the reverse shock \citep[see][]{Silich2018}.

\subsection{Impact of mass loading}
\label{sec4}

The distributions of the gas temperature and density within the hot blob around a representative 1~Myr 
old massive star located in the star cluster center are shown in Fig. 4. The cases without and with 
mass loading (dashed and solid lines, respectively) show very different results.
\begin{figure}
\includegraphics[width=\columnwidth]{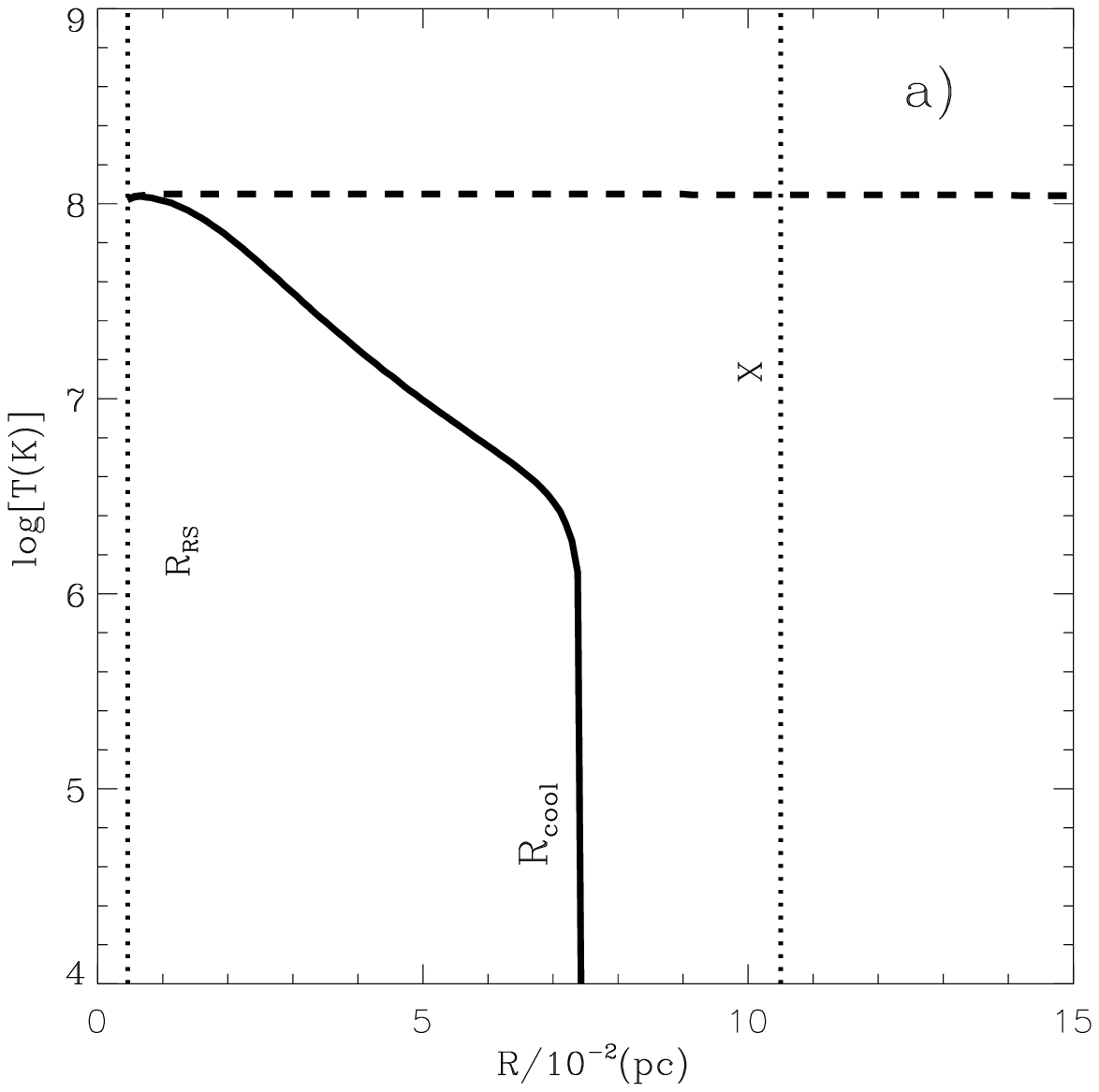}
\includegraphics[width=\columnwidth]{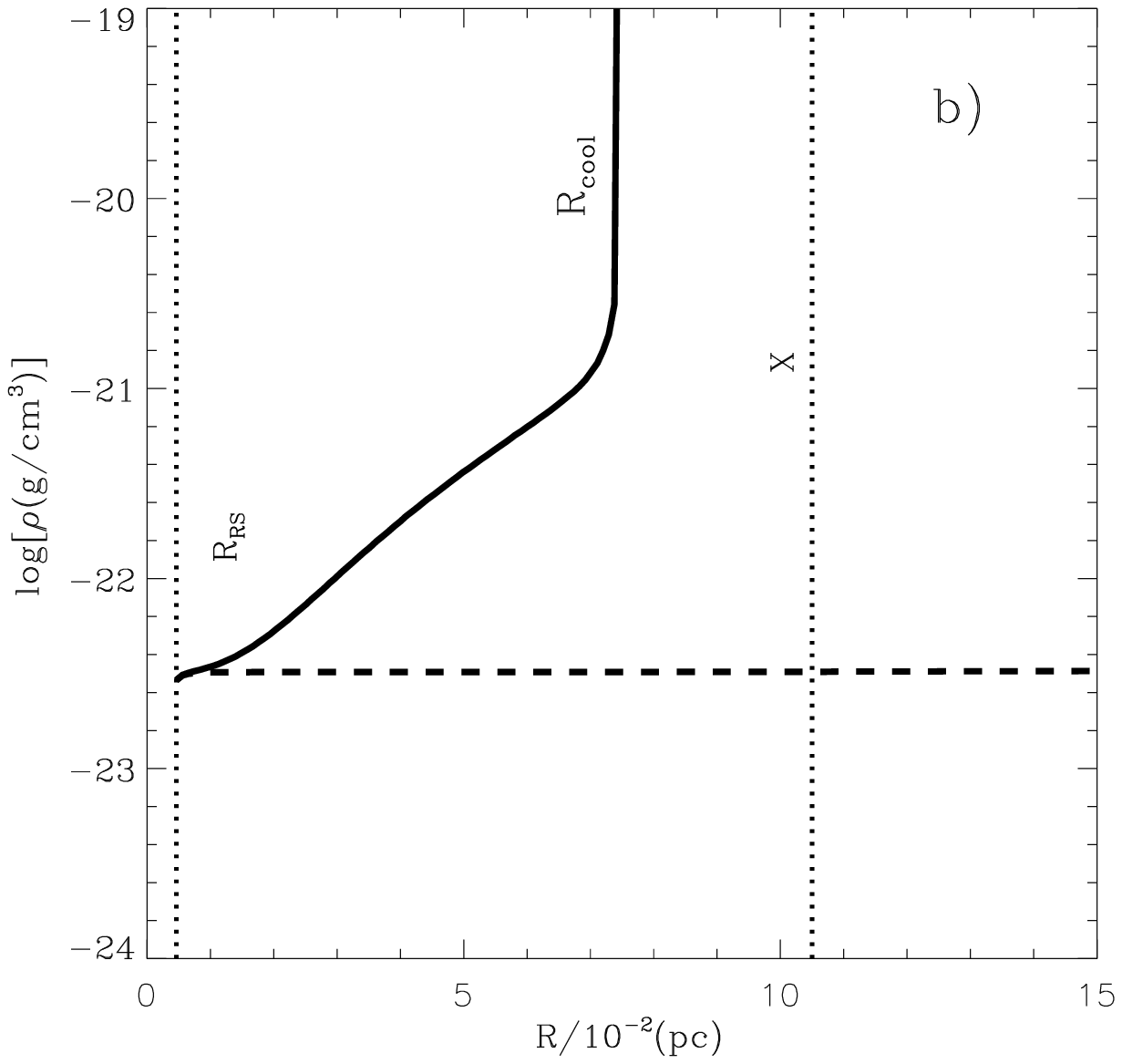}
\caption{The temperature and density distribution around a single massive star located in the star cluster 
center. Panels a and b present the temperature and density distributions in the case with and without 
mass loading (solid and dashed lines, respectively). Dotted vertical lines show, from the left to 
the right, the reverse shock position and the half distance between neighboring massive stars in the star 
cluster center, respectively. A mass loading rate of $5 \times 10^{-8}$\Msol \, yr$^{-1}$ per each PMS 
star was adopted in the mass loading model.}
\label{f4}
\end{figure}
Mass loading affects the temperature and density distributions around individual 
massive stars  significantly. In the model without mass loading the shocked wind gas does 
not cool and neighboring hot blobs merge to form a star cluster wind, whereas in the model 
with mass loading the shocked gas cools before merging at $R_{cool} < X$ which prevents the 
expulsion of the reinserted and the residual gas from the cluster. 

In order to determine if a star cluster could form a global wind or the global wind is suppressed, 
one should compare the cooling radii $R_{cool}$ with the half distance $X$ between neighboring 
massive stars. Fig. 5 shows how the $R_{cool} / X$ ratio changes with distance from a 1~Myr old 
star cluster center if the mass loading rate per each PMS star is  
${\dot M}_{PMS} = 5 \times 10^{-8}$\Msol \, yr$^{-1}$.
\begin{figure}
\includegraphics[width=\columnwidth]{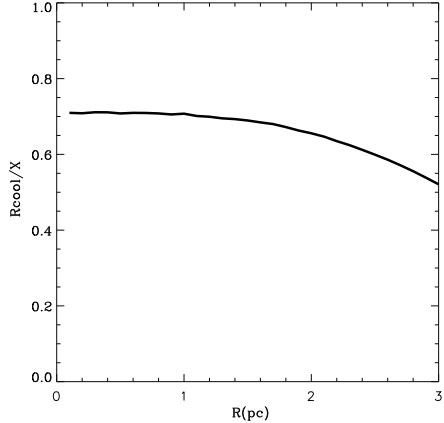}
\caption{The  $R_{cool} / X$ ratio as a function of the distance to the star cluster center.
              The same assumptions regarding the mass loading rate as those used in the previous section
              were adopted to calculate the $R_{cool}$ for massive stars located at different distances
              from the star cluster center.}  
 \label{f5}
\end{figure}
It is almost constant in the central zone of the cluster, where most massive stars are concentrated, and 
decreases at larger radii. This implies that if in the center $R_{cool} < X $, hot shocked blobs would 
not merge anywhere in the whole star cluster volume and the global star cluster wind would be 
suppressed. 

The shocked gas cooling radius evolution in the star cluster center is shown in Fig. 6. Here the cooling
radii are normalized to the half-distance between neighboring massive stars and compared for models 
with different mass loading rates. The first model (thick dashed line) assumes a conservative mass loading
rate of ${\dot M}_{PMS} = 8 \times 10^{-9}$\Msol \, yr$^{-1}$ per each pre-main sequence star
located inside the hot blob. About six times larger mass loading rate, ${\dot M}_{PMS} = 
5 \times 10^{-8}$\Msol \, yr$^{-1}$, was selected in the other case (thick solid line). In the 
case with a lower mass loading rate subsonic hot shocked winds merge and then form a  global star 
cluster wind. However, if the average mass loading rate per PMS star exceeds 
$8 \times 10^{-9}$\Msol \, yr$^{-1}$, hot shocked blobs may cool before merging. In 
this case the residual gas in the cluster remains for a while relatively undisturbed. 
The duration of this period is determined by the mass loading rate which in turn depends on the 
proto-stellar disks lifetime. The systematic survey for circumstellar disks in young stellar clusters 
indicates that the proto-stellar disk fraction decreases with the star cluster age and that about 
one-half of the stars within the cluster lose their disks within $\sim 3$~Myr 
\citep[see][]{Haisch2001,Lada2003,Fedele2010}. Therefore it is likely that in cloud D1 mass 
loading prevents the expulsion of the residual gas only for a few Myrs as during this time a 
significant number of proto-stellar disks vanishes and thus the mass loading rate in the shocked 
wind zones is reduced. This favors a suppressed negative stellar feedback scenario at a 
small cluster age and explains why the CO gas in this cluster remains relatively undisturbed.
\begin{figure}
\includegraphics[width=\columnwidth]{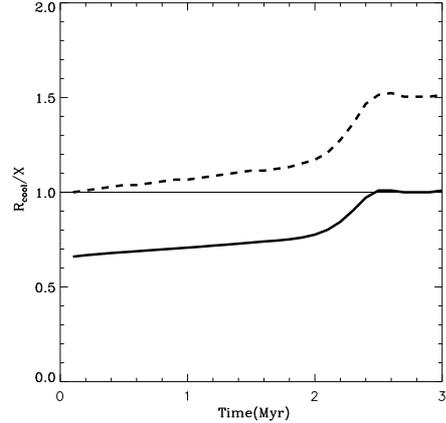}
\caption{Cooling radius $R_{cool}$ as a function of time. Dashed and solid lines present
cooling radii normalized to a half-distance between neighboring massive stars in the star cluster
center with ${\dot M}_{PMS} = 8 \times 10^{-9}$\Msol \, yr$^{-1}$ and
${\dot M}_{PMS} = 5 \times 10^{-8}$\Msol \, yr$^{-1}$ cases, respectively.}
 \label{f6}
\end{figure}

\subsection{Hot gas volume filling factor}

Calculating hot blobs cooling radii $R_{cool}(r)$ at different distances from the star cluster center, one 
can obtain the hot gas volume filling factor $f_X$:
\begin{equation}
      \label{eq11}
f_X = \frac{4 \pi}{R^3} \int_0^R  n_{mass}(r) R^3_{cool}(r) r^2 \dif{r} ,
\end{equation}
where $n_{mass}(r)$ is the massive star number density at different distances from the star cluster 
center and $R$ is the radius of the star cluster volume over which the filling factor is calculated. As 
massive stars are concentrated towards the center of cloud D1, $f_X$ depends on radius $R$.  $f_X$ 
calculated upon the assumption that ${\dot M}_{PMS} = 5 \times 10^{-8}$\Msol \, yr$^{-1}$ and  
$R = 1$~pc is shown in Fig. 7. There the filling factor $f_X$ in the central zone of the cluster grows 
slowly from $\sim 0.3$ at earliest stages of evolution to $\sim 0.5$ at the age of 2~Myr. It enhances 
then rapidly due to the increasing stellar wind power (see panel a on Fig. 2) to approach unity at the age 
of  about 2.4~Myr. Thus hot gas occupies a rather large fraction of the star cluster central zone even 
when a global star cluster wind is still suppressed. 
\begin{figure}
\includegraphics[width=\columnwidth]{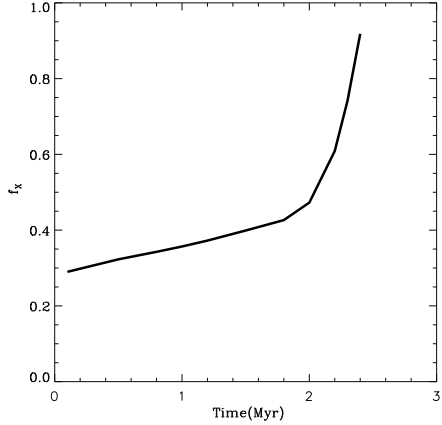}
\caption{The hot gas filling factor $f_X$ in the central zone of the cluster as a function of time.}
 \label{f7}
\end{figure}
\citet{Grimes2005} detected a diffuse X-ray emission from the NGC 5253 star-forming region; 
however at the resolution of Chandra, this emission cannot be definitely associated with the D1 
cluster. 

\subsection{The ionized gas distribution}

Before shocked winds merge, Lyman continuum photons escape from the shocked wind zones and
photoionize the residual gas between neighboring massive stars.  Dust grains and recombining atoms 
absorb these photons and form ionized shells around each massive star. The ionized gas density 
distribution in such shells and the shell mass are determined by the set of equations
\citep[see][]{Draine2011A,SergioMG2014}:   
\begin{eqnarray}
       \nonumber
      & & \hspace{-1.5cm} 
\frac{d}{dr}\left(\frac{\mu_a}{\mu_i} n k T_i + \mu_a n \sigma^2\right) = 
      \\[0.2cm]  \label{eq12a}
      & & \hspace{-0.9cm} 
n\sigma_d \frac{\left[L_n e^{-\tau}+L_i \phi\right]}{4\pi r^2 c} + 
n^2 \beta_2 \frac{\langle h\nu\rangle_i}{c} ,
      \\[0.2cm] \label{eq12b}
      & & \hspace{-1.5cm} 
\frac{d\phi}{dr} =  - \frac{4 \pi r^2 \beta_2 n^2}{Q_0}  - n \sigma_d \phi  ,
      \\[0.2cm] \label{eq12c}
      & & \hspace{-1.5cm}     
\frac{d\tau}{dr} = n\sigma_d ,
\end{eqnarray}
where $n(r)$ is the ionized gas density, $Q_0$ is the number of Lyman continuum photons emitted by 
the central star per second, $L_i$ and $L_n$ are the central star luminosities in ionizing and 
non-ionizing photons, respectively,  $\phi(r)$ is the fraction of the ionizing photons that reaches a surface 
with radius $r$ and $\langle h\nu\rangle_i = L_i/Q_0$ is the ionizing photons  mean energy.
 $k$ and $c$ are the Boltzmann constant and the speed of light,  $\beta_2 = 
2.59 \times 10^{-13}$~cm$^3$ s$^{-1}$ is the recombination coefficient to all but the ground level,
$\sigma_d$ is the effective dust absorption cross section per hydrogen atom, $\tau(r)$ is the dust 
optical depth and $\mu_a = 14/11 m_{H}$ is the mean mass per atom. 
$T_{i}= 10^4 \mbox{ K}$  is the ionized gas temperature and $\sigma$ is the residual gas velocity 
dispersion, which depends on the massive star position inside the cluster (see panel c in Fig.2).  Note 
that in addition to the ionized gas thermal pressure, equations (\ref{eq12a}-\ref{eq12c}) include the 
intra-cloud gas turbulent pressure. Following \citet{Draine2011A} we select the dust absorption cross 
section per hydrogen atom $\sigma_d =10^{-21}$ cm$^{2}$ as a standard value. Equations 
(\ref{eq12a}-\ref{eq12c}) are integrated 
from the cooling radius, where $\tau = 0$ and $\phi = 1$ outwards from the central star. The initial 
value of $n$ is selected from the condition that the ionized gas pressure at the inner edge of the
ionized shell  is equal to the gas thermal pressure inside a hot shocked wind blob.    
Starburst99 determines the central star mechanical power, $L_i$ and $L_n$  luminosities and the number 
of Lyman continuum photons $Q_0$. The location of a massive star with respect to the cluster center 
determines the shocked gas pressure and thus the value of the cooling radius and the ionized gas density 
at the inner edge $r = R_{cool}$ of the ionized zone. This allows one to determine the ionized zone 
thickness and mass around each massive star. 

The Lyman continuum photon flux and the gas density distribution around a 1 Myr old massive
star located in the star cluster center are shown in Fig. 8. Here a step in the gas density distribution
marks the edge of the HII region and is due to the different temperatures in the turbulent ionized 
($10^4$~K) and turbulent molecular \citep[$300$~K, see][]{Turner2015,Turner2017} gas 
components.
\begin{figure}
\includegraphics[width=\columnwidth]{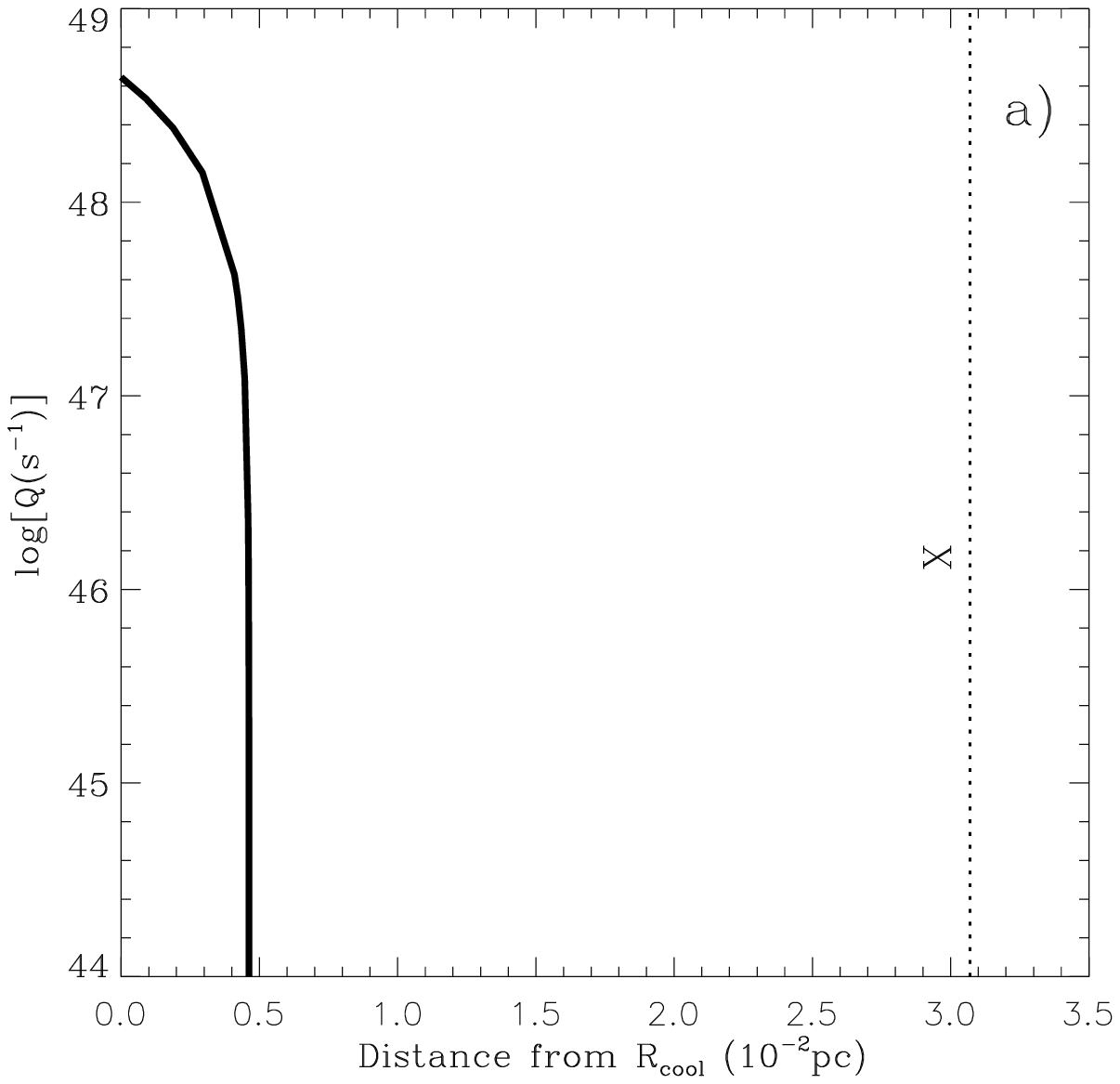}
\includegraphics[width=\columnwidth]{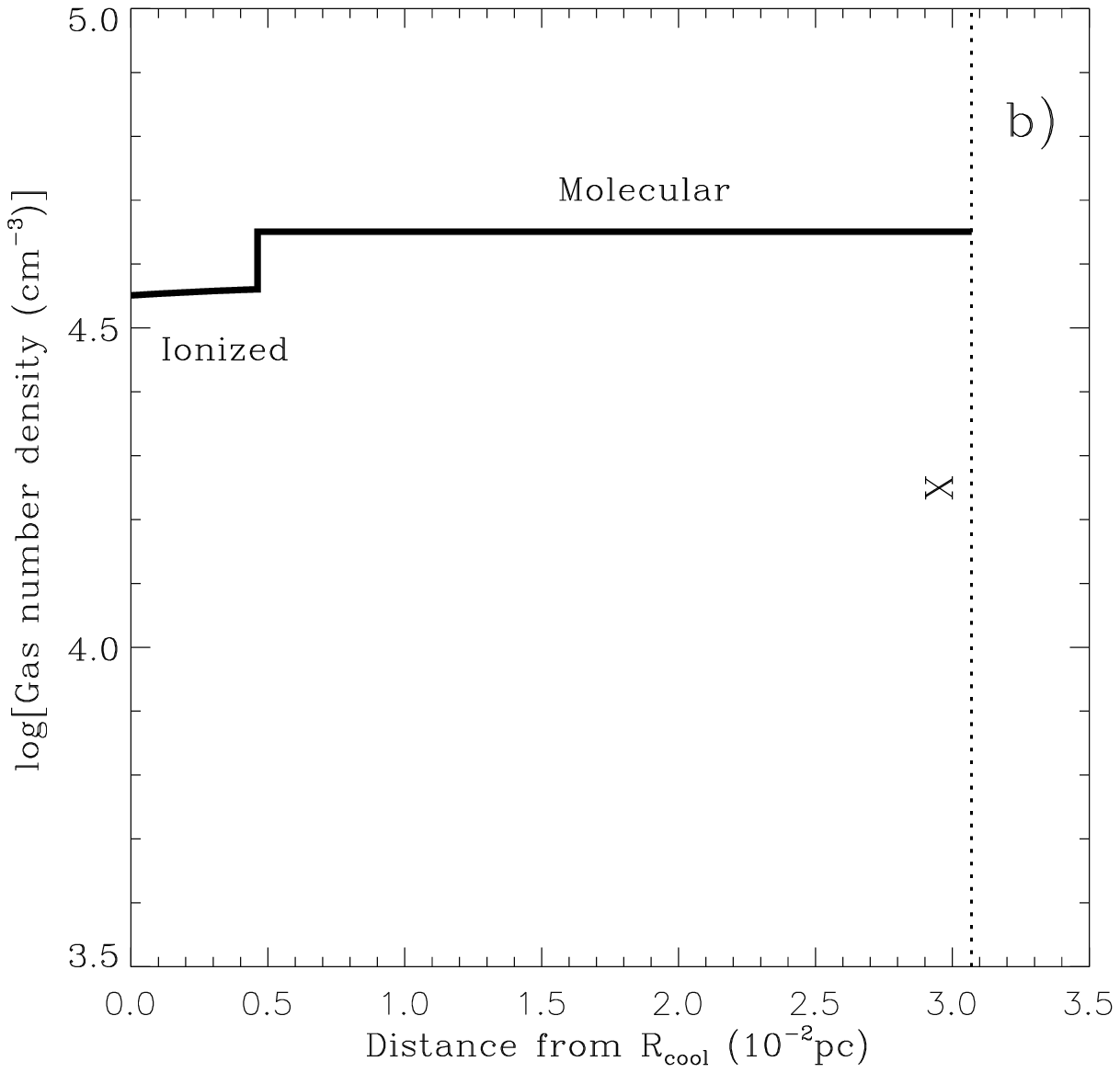}
\caption{Ultracompact HII regions around individual massive stars. Panel a shows the ionizing 
photon flux and panel b the ionized gas density distribution around a representative  1~Myr old 
massive star accommodated in the star cluster center. The vertical dotted line displays the 
half-distance to a neighboring massive star.} 
\label{f8}
\end{figure}
Note that the ionized shell is thin  and the ionized gas density is almost homogeneous, that 
implies a negligible impact of radiation pressure on the ionized shell structure. 

Having masses of ionized shells around massive stars located at different distances to the star cluster
center, one can obtain the total ionized gas mass $M_{HII}$:
\begin{equation}
      \label{eq13}
M_{HII} = 4 \pi \int_0^R  n_{mass}(r) m_{HII}(r) r^2 \dif{r} ,
\end{equation}
where $n_{mass}(r)$  and $m_{HII}(r)$ are the number of massive stars  per unit volume and the
ionized shell mass at different distances from the star cluster center and $R$ is the radius of the 
star cluster volume over which the ionized gas mass is calculated. The ionized gas mass inside a 3~pc 
central zone of the cluster at different times calculated upon the assumption that ${\dot M}_{PMS} = 
5 \times 10^{-8}$\Msol \, yr$^{-1}$ is shown in Fig. 9. 
\begin{figure}
\includegraphics[width=\columnwidth]{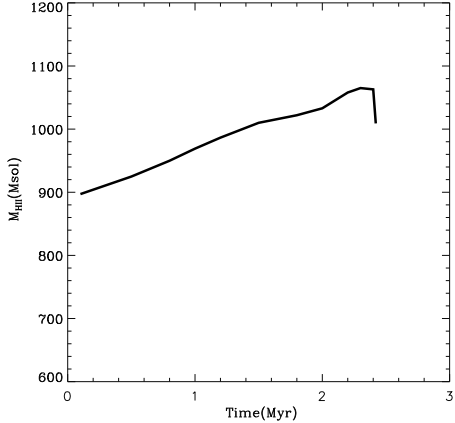}
\caption{The ionized intra-cloud gas mass as a function of time. The ionized gas mass inside a 3~pc 
central zone of D1 cloud was calculated upon the assumption that the mass-loss rate per each PMS star 
is $5 \times 10^{-8}$\Msol \, yr$^{-1}$.}
 \label{f9}
\end{figure}
The ionized gas mass in the central zone of the cluster increases slowly until hot shocked winds 
start to merge. The ionized gas mass drops then as no neutral/molecular gas remains in the central zone 
of the cluster. Lyman continuum photons then escape from the central zone to ionize gas located at 
larger distances from the cloud center. Note that photoionization of proto-stellar disks depends 
on many factors, such as the disk size, orientation, distance to the massive star and was not
considered in these calculations.

D1 cluster is a very strong source of the infrared emission \citep{Gorjian2001} whose spectral
energy distribution (SED) presents a clear near-infrared excess \citep{AlonsoHerrero2004}. Here we
asssume that ionized shells around massive stars are dusty and make use the theory of 
stochastic grain temperature fluctuations \citep{Guhathakurta1989,Sergio2016,Sergio2017} to
obtain the infrared SED associated with these dusty ultracompact HII regions located in the central
zone of the cluster. The model predicted cooling radii and the ionized shell masses together with 
the representative 1~Myr old massive star Starburst99 spectrum and the gas-to-dust mass ratio
obtained by \citet{Turner2015} (GTD=47) were used as the input parameters for these calculations.
It was also assumed that dust grains have a power-low size distribution $\sim a^{-3.5}$ with the lower 
and upper cutoff radii $a_{min}=0.001\mu$m and $a_{max}=0.5\mu$m, respectively, that carbonaceous
and silicate grains have an equal mass proportion. 

The integrated over a star cluster volume with radius $R$ infrared SED then is:
\begin{equation}
      \label{eq14}
F_{\nu} = 4 \pi \int_0^R  n_{mass}(r) f_{\nu}(r) r^2 \dif{r} ,
\end{equation}
where $n_{mass}(r)$ and $f_{\nu}(r)$ are the number of massive stars  per unit volume and the spectral 
energy distribution associated with individual massive stars located at different distances from the star 
cluster center, respectively. $F_{\nu}$ is the integrated over the star cluster volume SED. Figure 
10 presents $F_{\nu}$ that accounts for all massive stars located within cloud D1 central zone with
radius $R = 3$pc. It agrees well with the near-IR excess revealed by \citet{Vanzi2004}, which is 
displayed in Fig. 10 by the triangle symbols.
\begin{figure}
\includegraphics[width=\columnwidth]{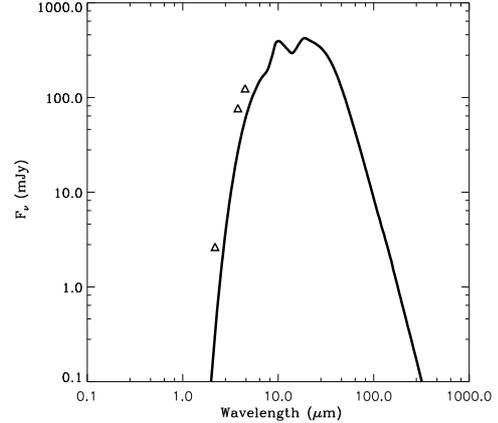}
\caption{The model-predicted IR spectral energy distribution. Triangle symbols display the near-IR
excess revealed by \citet{Vanzi2004}.} 
 \label{f10}
\end{figure}

\subsection{Model Uncertainties and Simplifications}

In the previous sections we examined the interplay among massive star winds and the
residual gas in massive star-forming molecular clouds at scales comparable to the mean
separation between massive stars in the assembling cluster.  Several significant simplifications
were adopted to reach our major aim - to reveal physical processes which may suppress or delay
a global star cluster wind development and the dispersal of the parental molecular cloud at the
early, pre-supernovae, star cluster formation stage. In particular, we adopted a static distribution
of stars and gas in the parental cloud and thus considered a snapshot of a more 
complicated dynamical process \citep[e.g.][]{Lahen2019A,Lahen2019B}. We also did not
consider a possible mass segregation and thus a possible difference in massive and low mass
star distributions. Another significant simplification is the assumption that all massive stars
are identical whereas mechanical power and mass loss rates strongly depend upon the
star mass that may lead to the formation of a more complicated structure with less massive
star winds enclosed into the pockets of larger hot shocked winds formed around most massive 
stars. We were also unable to calculate how mass loading rate from PMS stars change with time 
within young massive clusters and therefore derived a lower mass loading rate per PMS star which is
required to suppress the development of a global star cluster wind and delay the residual gas 
expulsion from the star cluster parental cloud D1. Finally,
a spherical approximation used in the calculations is valid until shocked wind cooling radii
remain smaller than the characteristic scale (core radius) of the residual gas distribution. This
approximation is good for this particular cluster as in this case massive stars are located in the 
central zone of dense D1 cloud where the residual gas distribution is almost homogeneous and the 
calculated shocked gas cooling radii are much smaller than the molecular cloud core radius. 
However, it may be violated in systems with similar characteristic scales for the residual gas and massive 
star distributions. Further understanding of early star cluster formation stages certainly would benefit 
from the detailed consideration of all these effects.

\section{Summary and Conclusions}

Here the interplay among massive stars and the residual gas in young stellar clusters was
investigated. It was suggested that mass loading from PMS stars may have a  significant
impact on the early stages of the star cluster assembling. Mass loading slows down the flow and 
enhances the cooling rate in the hot shocked wind zones around massive stars that may suppress 
or delay a global star 
cluster wind development and the dispersal of the parental molecular cloud at the pre-supernovae
star cluster formation stage. This effect was incorporated into the model developed in our previous 
papers \citep{Silich2017, Silich2018} and confronted with recent observations of an obscured 
young stellar cluster in the center of NGC 5253 D1 molecular cloud.

Following radio, IR and CO observations, it was adopted that in the NGC 5253 D1 cluster recently formed 
stars are highly concentrated towards the cloud D1 center and the characteristic scales for the molecular 
gas and stellar mass distributions are different: $\approx 2.4$~pc and $\approx 0.8$~pc for the molecular 
gas and stars, respectively. In this case the central gas density and the central turbulent pressure are large
and exceed $\sim 10^4$~cm$^{-3}$ and  $10^{-7}$~dyn cm$^{-2}$, respectively. These parameters
together with the molecular cloud mass and the star formation efficiency have been used as input 
parameters for simulations. 

The calculations without mass loading show that despite the turbulent pressure in the central 
zone of such molecular cloud is large, it cannot prevent neighboring hot shocked winds from 
merging. This is in conflict with ALMA observations which show that the CO gas in cloud D1 is 
relatively undisturbed. However, in the calculations with a moderate mass loading rate 
stellar winds do not merge at least during a few Myr.  This delays the development of a global 
star cluster wind and the progenetor molecular cloud dispersal.  The negative stellar feedback 
is then suppressed and the residual molecular gas remains relatively undisturbed.

These simulations show that in the case of NGC 5253 D1 cloud mass loading rates larger than 
${\dot M}_{PMS} = 8 \times 10^{-9}$\Msol \, yr$^{-1}$ per each low mass PMS star are
required to suppress the negative stellar feedback on the parental molecular cloud.
The duration of the suppressed negative feedback stage depends on the mass loading rate, 
which ceases with time, and massive star mechanical luminosities, which increase with 
time. In the D1 cluster it may reach a few Myr. During this stage the star-forming 
cloud remains relatively undisturbed. Hot shocked winds cannot merge and occupy only a fraction 
of cloud D1 central zone. In our representative model with a constant mass loading rate of
$5 \times 10^{-8}$\Msol \, yr$^{-1}$ per each PMS low mass star, the hot component volume 
filling factor in the central 1~pc zone of the cloud is moderate ($f_X \approx 0.3$) at the earliest 
stages of evolution. However, it grows to unity at the age of $\approx 2.4$~Myr due to increasing 
power of individual stellar winds. 

In contrast, the ionized residual gas is concentrated in very thin, high density layers around 
massive stars 
and occupies only a small fraction of D1 cloud volume. The model predicted ionized gas 
mass in the central zone of the cloud ($\sim 1000$\Msol) is smaller than estimated by
\citet{Turner2004}  ($\sim 1900$\Msol). However, the last estimate was obtained under the
assumption of the ionized gas unity filling factor. A smaller filling factor would enhance the 
ionized gas density that, in turn, would result in a lower ionized gas mass. We thus conclude that
the model predicted ionized gas mass and the immersed dust spectral energy distribution are in
good agreement with the radio \citep{Turner2004} and near-IR \citet{Vanzi2004} 
observations of cloud D1 and its embedded cluster.

The model explains how the impact of stellar winds on the residual molecular gas  or the 
negative stellar feedback in NGC 5253 D1 cluster could be suppressed.
However, it is unlikely that photo-evaporation of proto-stellar disks may last for a long time
even at the required moderate level as the number of proto-stellar disks decreases with the 
star cluster age \citep{Johnstone1998,Storzer1999,Richling2000}. In the case of NGC 5253 D1 
cloud this favors a short (a few Myrs) suppressed negative feedback scenario for the embedded 
cluster. It is expected that after this time hot shocked winds around individual massive stars 
merge and form a global star cluster wind.

\appendix

\section{Pre-main sequence stars in young stellar clusters}
  
Pre-main sequence low mass stars may contribute significantly to the mass balance in young stellar 
clusters. Starburst99 model does not include pre-main sequence stars contribution. Therefore here the 
number of low mass ($M < 3$\Msol) stars in a stellar cluster with a standard Kroupa IMF is determined. 
This number is used then to estimate the mass balance in hot shock-heated zones around individual 
massive stars. The number of stars per unit mass interval in the cluster with a broken power law IMF and
lower and upper mass cutoffs $M_{low}$ and $M_{up}$ is:     
\begin{eqnarray}
     \label{eqA1}
      & & \hspace{-1.5cm}
N_1(m) = A m^{-\alpha_1} \qquad m \le m_1 ,
       \\[0.2cm]  \label{eqA2}
     & & \hspace{-1.5cm}
N_2(m) =  B m^{-\alpha_2} \qquad m \ge m_1 ,
\end{eqnarray}
where $m_1$ is the turnoff mass. The requirement for the IMF to be continuous at $m_1$ yields:
\begin{equation}
       \label{eqA3}
A = B m_1^{\alpha_1 - \alpha_2} .
\end{equation}
The total number of stars with masses smaller than $M$ is:
\begin{eqnarray}
         \nonumber
      & & \hspace{-1.5cm}
N(m < M) = \frac{B}{1-\alpha_1} \left[\left(m_1^{1-\alpha_2} - 
                     \frac{M_{low}^{1-\alpha_1}}{m_1^{\alpha_2-\alpha_1}}\right) \quad + \right.
       \\[0.2cm]  \label{eqA4}
     & & \hspace{-0.0cm}
     \left. \frac{1-\alpha_1}{1-\alpha_2} \left(M^{1-\alpha_2} - m_1^{1-\alpha_2}\right)\right] .
\end{eqnarray}
The normalization coefficient $B$ in equation (\ref{eqA4}) is determined by the total mass 
of the cluster $M_{\star}$:
\begin{equation}
       \label{eqA5}
B = \frac{(2-\alpha_1) M_{\star}}{m_1^{2-\alpha_2} - 
        \frac{M_{low}^{2-\alpha_1}}{m_1^{\alpha_2-\alpha_1}} + \frac{2-\alpha_1}{2-\alpha_2}
        \left(M_{up}^{2-\alpha_2} - m_1^{2-\alpha_2}\right)} .
\end{equation}
In the case of a  standard Kroupa IMF $M_{low} = 0.1$\Msol,  $M_1 = 0.5$\Msol, $\alpha_1 = 1.3$ and
$\alpha_2 = 2.3$ \citep[e.g.][]{Leitherer1999}. PMS low mass stars have masses $M < 3$\Msol \,
\citep[e.g.][]{Appenzeller1989}. Therefore the initial number of low mass PMS stars in a young stellar 
cluster with a Standard Kroupa IMF and an upper cutoff mass 100\Msol \, scales with the star cluster 
mass as:
\begin{equation}
      \label{eqA6}
N_{PMS} \approx 1.5 \times 10^6 (M_{\star}/10^6\Msol) .
\end{equation}

\section*{Acknowledgements}

We thank the anonymous referee for many suggestions which substantially improved the original version 
of the paper. This study was supported by CONACYT México, research grant A1-S-28458.  S.M.G. also 
acknowledges support of CONACYT through C\'atedra n.482. JT acknowledges support of NSF 
through grant AST1515570. The authors also acknowledge the support provided by the Laboratorio 
Nacional de Superc\'omputo del Sureste de M\'exico, CONACYT member of the national laboratories
network and thank the participants of the 2019 Guillermo Haro Workshop, in particular Linda Smith, 
Sara Beck and Michelle Consiglio, for helpful discussions.

\bibliographystyle{mnras}
\bibliography{GC}

\begin{thebibliography}{}
\makeatletter
\relax
\def\mn@urlcharsother{\let\do\@makeother \do\$\do\&\do\#\do\^\do\_\do\%\do\~}
\def\mn@doi{\begingroup\mn@urlcharsother \@ifnextchar [ {\mn@doi@}
  {\mn@doi@[]}}
\def\mn@doi@[#1]#2{\def\@tempa{#1}\ifx\@tempa\@empty \href
  {http://dx.doi.org/#2} {doi:#2}\else \href {http://dx.doi.org/#2} {#1}\fi
  \endgroup}
\def\mn@eprint#1#2{\mn@eprint@#1:#2::\@nil}
\def\mn@eprint@arXiv#1{\href {http://arxiv.org/abs/#1} {{\tt arXiv:#1}}}
\def\mn@eprint@dblp#1{\href {http://dblp.uni-trier.de/rec/bibtex/#1.xml}
  {dblp:#1}}
\def\mn@eprint@#1:#2:#3:#4\@nil{\def\@tempa {#1}\def\@tempb {#2}\def\@tempc
  {#3}\ifx \@tempc \@empty \let \@tempc \@tempb \let \@tempb \@tempa \fi \ifx
  \@tempb \@empty \def\@tempb {arXiv}\fi \@ifundefined
  {mn@eprint@\@tempb}{\@tempb:\@tempc}{\expandafter \expandafter \csname
  mn@eprint@\@tempb\endcsname \expandafter{\@tempc}}}

\bibitem[\protect\citeauthoryear{{Adams}, {Kochanek}, {Gerke}, {Stanek}  \&
  {Dai}}{{Adams} et~al.}{2017a}]{Adams2017A}
{Adams} S.~M.,  {Kochanek} C.~S.,  {Gerke} J.~R.,  {Stanek} K.~Z.,   {Dai} X.,
  2017a, \mn@doi [\mnras] {10.1093/mnras/stx816}, \href
  {http://adsabs.harvard.edu/abs/2017MNRAS.468.4968A} {468, 4968}

\bibitem[\protect\citeauthoryear{{Adams}, {Kochanek}, {Gerke}  \&
  {Stanek}}{{Adams} et~al.}{2017b}]{Adams2017B}
{Adams} S.~M.,  {Kochanek} C.~S.,  {Gerke} J.~R.,   {Stanek} K.~Z.,  2017b,
  \mn@doi [\mnras] {10.1093/mnras/stx898}, \href
  {http://adsabs.harvard.edu/abs/2017MNRAS.469.1445A} {469, 1445}

\bibitem[\protect\citeauthoryear{{Alonso-Herrero}, {Takagi}, {Baker}, {Rieke},
  {Rieke}, {Imanishi}  \& {Scoville}}{{Alonso-Herrero}
  et~al.}{2004}]{AlonsoHerrero2004}
{Alonso-Herrero} A.,  {Takagi} T.,  {Baker} A.~J.,  {Rieke} G.~H.,  {Rieke}
  M.~J.,  {Imanishi} M.,   {Scoville} N.~Z.,  2004, \mn@doi [\apj]
  {10.1086/422448}, \href
  {https://ui.adsabs.harvard.edu/abs/2004ApJ...612..222A} {612, 222}

\bibitem[\protect\citeauthoryear{{Appenzeller} \& {Mundt}}{{Appenzeller} \&
  {Mundt}}{1989}]{Appenzeller1989}
{Appenzeller} I.,  {Mundt} R.,  1989, \mn@doi [\aapr] {10.1007/BF00873081},
  \href {https://ui.adsabs.harvard.edu/abs/1989A%26ARv...1..291A} {1, 291}

\bibitem[\protect\citeauthoryear{{Bally}, {Reipurth}  \& {Davis}}{{Bally}
  et~al.}{2007}]{Bally2007}
{Bally} J.,  {Reipurth} B.,   {Davis} C.~J.,  2007, Protostars and Planets V,
  \href {https://ui.adsabs.harvard.edu/abs/2007prpl.conf..215B} {pp 215--230}

\bibitem[\protect\citeauthoryear{{Banerjee} \& {Kroupa}}{{Banerjee} \&
  {Kroupa}}{2015}]{Banerjee2015}
{Banerjee} S.,  {Kroupa} P.,  2015, \mn@doi [\mnras] {10.1093/mnras/stu2445},
  \href {https://ui.adsabs.harvard.edu/abs/2015MNRAS.447..728B} {447, 728}

\bibitem[\protect\citeauthoryear{{Bastian} \& {Lardo}}{{Bastian} \&
  {Lardo}}{2015}]{Bastian2015B}
{Bastian} N.,  {Lardo} C.,  2015, \mn@doi [\mnras] {10.1093/mnras/stv1661},
  \href {http://adsabs.harvard.edu/abs/2015MNRAS.453..357B} {453, 357}

\bibitem[\protect\citeauthoryear{{Bastian}, {Hollyhead}  \&
  {Cabrera-Ziri}}{{Bastian} et~al.}{2014}]{Bastian2014}
{Bastian} N.,  {Hollyhead} K.,   {Cabrera-Ziri} I.,  2014, \mn@doi [\mnras]
  {10.1093/mnras/stu1775}, \href
  {http://adsabs.harvard.edu/abs/2014MNRAS.445..378B} {445, 378}

\bibitem[\protect\citeauthoryear{{Baumgardt} \& {Kroupa}}{{Baumgardt} \&
  {Kroupa}}{2007}]{Baumgardt2007}
{Baumgardt} H.,  {Kroupa} P.,  2007, \mn@doi [\mnras]
  {10.1111/j.1365-2966.2007.12209.x}, \href
  {http://adsabs.harvard.edu/abs/2007MNRAS.380.1589B} {380, 1589}

\bibitem[\protect\citeauthoryear{{Beck}}{{Beck}}{2015}]{Beck2015}
{Beck} S.,  2015, \mn@doi [International Journal of Modern Physics D]
  {10.1142/S0218271815300025}, \href
  {http://adsabs.harvard.edu/abs/2015IJMPD..2430002B} {24, 1530002}

\bibitem[\protect\citeauthoryear{{Beck}, {Turner}, {Ho}, {Lacy}  \&
  {Kelly}}{{Beck} et~al.}{1996}]{Beck1996}
{Beck} S.~C.,  {Turner} J.~L.,  {Ho} P.~T.~P.,  {Lacy} J.~H.,   {Kelly} D.~M.,
  1996, \mn@doi [\apj] {10.1086/176757}, \href
  {http://adsabs.harvard.edu/abs/1996ApJ...457..610B} {457, 610}

\bibitem[\protect\citeauthoryear{{Bedin}, {Piotto}, {Anderson}, {Cassisi},
  {King}, {Momany}  \& {Carraro}}{{Bedin} et~al.}{2004}]{Bedin2004}
{Bedin} L.~R.,  {Piotto} G.,  {Anderson} J.,  {Cassisi} S.,  {King} I.~R.,
  {Momany} Y.,   {Carraro} G.,  2004, \mn@doi [\apjl] {10.1086/420847}, \href
  {http://adsabs.harvard.edu/abs/2004ApJ...605L.125B} {605, L125}

\bibitem[\protect\citeauthoryear{{Boily} \& {Kroupa}}{{Boily} \&
  {Kroupa}}{2003}]{Boily2003}
{Boily} C.~M.,  {Kroupa} P.,  2003, \mn@doi [\mnras]
  {10.1046/j.1365-8711.2003.06076.x}, \href
  {http://adsabs.harvard.edu/abs/2003MNRAS.338..665B} {338, 665}

\bibitem[\protect\citeauthoryear{{Calura}, {Few}, {Romano}  \&
  {D'Ercole}}{{Calura} et~al.}{2015}]{Calura2015}
{Calura} F.,  {Few} C.~G.,  {Romano} D.,   {D'Ercole} A.,  2015, \mn@doi
  [\apjl] {10.1088/2041-8205/814/1/L14}, \href
  {http://adsabs.harvard.edu/abs/2015ApJ...814L..14C} {814, L14}

\bibitem[\protect\citeauthoryear{{Calzetti} et~al.,}{{Calzetti}
  et~al.}{2015}]{Calzetti2015}
{Calzetti} D.,  et~al., 2015, \mn@doi [\apj] {10.1088/0004-637X/811/2/75},
  \href {http://adsabs.harvard.edu/abs/2015ApJ...811...75C} {811, 75}

\bibitem[\protect\citeauthoryear{{Carretta} et~al.,}{{Carretta}
  et~al.}{2009}]{Carretta2009}
{Carretta} E.,  et~al., 2009, \mn@doi [\aap] {10.1051/0004-6361/200912096},
  \href {http://adsabs.harvard.edu/abs/2009A%26A...505..117C} {505, 117}

\bibitem[\protect\citeauthoryear{{Coffey}, {Bacciotti}  \& {Podio}}{{Coffey}
  et~al.}{2008}]{Coffey2008}
{Coffey} D.,  {Bacciotti} F.,   {Podio} L.,  2008, \mn@doi [\apj]
  {10.1086/592343}, \href
  {https://ui.adsabs.harvard.edu/abs/2008ApJ...689.1112C} {689, 1112}

\bibitem[\protect\citeauthoryear{{Cohen}, {Turner}, {Consiglio}, {Martin}  \&
  {Beck}}{{Cohen} et~al.}{2018}]{Cohen2018}
{Cohen} D.~P.,  {Turner} J.~L.,  {Consiglio} S.~M.,  {Martin} E.~C.,   {Beck}
  S.~C.,  2018, \mn@doi [\apj] {10.3847/1538-4357/aac170}, \href
  {https://ui.adsabs.harvard.edu/abs/2018ApJ...860...47C} {860, 47}

\bibitem[\protect\citeauthoryear{{Dale}, {Ercolano}  \& {Bonnell}}{{Dale}
  et~al.}{2015}]{Dale2015}
{Dale} J.~E.,  {Ercolano} B.,   {Bonnell} I.~A.,  2015, \mn@doi [\mnras]
  {10.1093/mnras/stv913}, \href
  {https://ui.adsabs.harvard.edu/abs/2015MNRAS.451..987D} {451, 987}

\bibitem[\protect\citeauthoryear{{Draine}}{{Draine}}{2011}]{Draine2011A}
{Draine} B.~T.,  2011, \mn@doi [\apj] {10.1088/0004-637X/732/2/100}, \href
  {http://adsabs.harvard.edu/abs/2011ApJ...732..100D} {732, 100}

\bibitem[\protect\citeauthoryear{{Drake}, {Ercolano}, {Flaccomio}  \&
  {Micela}}{{Drake} et~al.}{2009}]{Drake2009}
{Drake} J.~J.,  {Ercolano} B.,  {Flaccomio} E.,   {Micela} G.,  2009, \mn@doi
  [\apjl] {10.1088/0004-637X/699/1/L35}, \href
  {https://ui.adsabs.harvard.edu/abs/2009ApJ...699L..35D} {699, L35}

\bibitem[\protect\citeauthoryear{{Elmegreen}}{{Elmegreen}}{2017}]{Elmegreen2017}
{Elmegreen} B.~G.,  2017, \mn@doi [\apj] {10.3847/1538-4357/836/1/80}, \href
  {http://adsabs.harvard.edu/abs/2017ApJ...836...80E} {836, 80}

\bibitem[\protect\citeauthoryear{{Elmegreen} \& {Efremov}}{{Elmegreen} \&
  {Efremov}}{1997}]{Elmegreen1997}
{Elmegreen} B.~G.,  {Efremov} Y.~N.,  1997, \mn@doi [\apj] {10.1086/303966},
  \href {http://adsabs.harvard.edu/abs/1997ApJ...480..235E} {480, 235}

\bibitem[\protect\citeauthoryear{{Ercolano}, {Drake}, {Raymond}  \&
  {Clarke}}{{Ercolano} et~al.}{2008}]{Ercolano2008}
{Ercolano} B.,  {Drake} J.~J.,  {Raymond} J.~C.,   {Clarke} C.~C.,  2008,
  \mn@doi [\apj] {10.1086/590490}, \href
  {https://ui.adsabs.harvard.edu/abs/2008ApJ...688..398E} {688, 398}

\bibitem[\protect\citeauthoryear{{Fedele}, {van den Ancker}, {Henning},
  {Jayawardhana}  \& {Oliveira}}{{Fedele} et~al.}{2010}]{Fedele2010}
{Fedele} D.,  {van den Ancker} M.~E.,  {Henning} T.,  {Jayawardhana} R.,
  {Oliveira} J.~M.,  2010, \mn@doi [\aap] {10.1051/0004-6361/200912810}, \href
  {https://ui.adsabs.harvard.edu/abs/2010A%26A...510A..72F} {510, A72}

\bibitem[\protect\citeauthoryear{{Finn}, {Johnson}, {Brogan}, {Wilson},
  {Indebetouw}, {Harris}, {Kamenetzky}  \& {Bemis}}{{Finn}
  et~al.}{2019}]{Finn2019}
{Finn} M.~K.,  {Johnson} K.~E.,  {Brogan} C.~L.,  {Wilson} C.~D.,  {Indebetouw}
  R.,  {Harris} W.~E.,  {Kamenetzky} J.,   {Bemis} A.,  2019, \mn@doi [\apj]
  {10.3847/1538-4357/ab0d1e}, \href
  {https://ui.adsabs.harvard.edu/abs/2019ApJ...874..120F} {874, 120}

\bibitem[\protect\citeauthoryear{{Frank} et~al.,}{{Frank}
  et~al.}{2014}]{Frank2014}
{Frank} A.,  et~al., 2014, \mn@doi [Protostars and Planets VI]
  {10.2458/azu_uapress_9780816531240-ch020}, \href
  {https://ui.adsabs.harvard.edu/abs/2014prpl.conf..451F} {pp 451--474}

\bibitem[\protect\citeauthoryear{{Gorjian}, {Turner}  \& {Beck}}{{Gorjian}
  et~al.}{2001}]{Gorjian2001}
{Gorjian} V.,  {Turner} J.~L.,   {Beck} S.~C.,  2001, \mn@doi [\apjl]
  {10.1086/320923}, \href {http://adsabs.harvard.edu/abs/2001ApJ...554L..29G}
  {554, L29}

\bibitem[\protect\citeauthoryear{{Grimes}, {Heckman}, {Strickland}  \&
  {Ptak}}{{Grimes} et~al.}{2005}]{Grimes2005}
{Grimes} J.~P.,  {Heckman} T.,  {Strickland} D.,   {Ptak} A.,  2005, \mn@doi
  [\apj] {10.1086/430692}, \href
  {https://ui.adsabs.harvard.edu/abs/2005ApJ...628..187G} {628, 187}

\bibitem[\protect\citeauthoryear{{Guhathakurta} \& {Draine}}{{Guhathakurta} \&
  {Draine}}{1989}]{Guhathakurta1989}
{Guhathakurta} P.,  {Draine} B.~T.,  1989, \mn@doi [\apj] {10.1086/167899},
  \href {https://ui.adsabs.harvard.edu/abs/1989ApJ...345..230G} {345, 230}

\bibitem[\protect\citeauthoryear{{Haisch}, {Lada}  \& {Lada}}{{Haisch}
  et~al.}{2001}]{Haisch2001}
{Haisch} Jr. K.~E.,  {Lada} E.~A.,   {Lada} C.~J.,  2001, \mn@doi [\apjl]
  {10.1086/320685}, \href
  {https://ui.adsabs.harvard.edu/abs/2001ApJ...553L.153H} {553, L153}

\bibitem[\protect\citeauthoryear{{Hartmann} \& {Bae}}{{Hartmann} \&
  {Bae}}{2018}]{Hartmann2018}
{Hartmann} L.,  {Bae} J.,  2018, \mn@doi [\mnras] {10.1093/mnras/stx2775},
  \href {https://ui.adsabs.harvard.edu/abs/2018MNRAS.474...88H} {474, 88}

\bibitem[\protect\citeauthoryear{{Johnson}, {Leroy}, {Indebetouw}, {Brogan},
  {Whitmore}, {Hibbard}, {Sheth}  \& {Evans}}{{Johnson}
  et~al.}{2015}]{Johnson2015}
{Johnson} K.~E.,  {Leroy} A.~K.,  {Indebetouw} R.,  {Brogan} C.~L.,  {Whitmore}
  B.~C.,  {Hibbard} J.,  {Sheth} K.,   {Evans} A.~S.,  2015, \mn@doi [\apj]
  {10.1088/0004-637X/806/1/35}, \href
  {http://adsabs.harvard.edu/abs/2015ApJ...806...35J} {806, 35}

\bibitem[\protect\citeauthoryear{{Johnstone}, {Hollenbach}  \&
  {Bally}}{{Johnstone} et~al.}{1998}]{Johnstone1998}
{Johnstone} D.,  {Hollenbach} D.,   {Bally} J.,  1998, \mn@doi [\apj]
  {10.1086/305658}, \href
  {https://ui.adsabs.harvard.edu/abs/1998ApJ...499..758J} {499, 758}

\bibitem[\protect\citeauthoryear{{Kim} \& {Lee}}{{Kim} \&
  {Lee}}{2018}]{Kim2018}
{Kim} J.~J.,  {Lee} Y.-W.,  2018, \mn@doi [\apj] {10.3847/1538-4357/aaec67},
  \href {http://adsabs.harvard.edu/abs/2018ApJ...869...35K} {869, 35}

\bibitem[\protect\citeauthoryear{{Krause}, {Charbonnel}, {Decressin}, {Meynet},
  {Prantzos}  \& {Diehl}}{{Krause} et~al.}{2012}]{Krause2012}
{Krause} M.,  {Charbonnel} C.,  {Decressin} T.,  {Meynet} G.,  {Prantzos} N.,
  {Diehl} R.,  2012, \mn@doi [\aap] {10.1051/0004-6361/201220244}, \href
  {http://adsabs.harvard.edu/abs/2012A%26A...546L...5K} {546, L5}

\bibitem[\protect\citeauthoryear{{Krause}, {Charbonnel}, {Bastian}  \&
  {Diehl}}{{Krause} et~al.}{2016}]{Krause2016}
{Krause} M.~G.~H.,  {Charbonnel} C.,  {Bastian} N.,   {Diehl} R.,  2016,
  \mn@doi [\aap] {10.1051/0004-6361/201526685}, \href
  {http://adsabs.harvard.edu/abs/2016A%26A...587A..53K} {587, A53}

\bibitem[\protect\citeauthoryear{{Lada} \& {Lada}}{{Lada} \&
  {Lada}}{2003}]{Lada2003}
{Lada} C.~J.,  {Lada} E.~A.,  2003, \mn@doi [\araa]
  {10.1146/annurev.astro.41.011802.094844}, \href
  {https://ui.adsabs.harvard.edu/abs/2003ARA%26A..41...57L} {41, 57}

\bibitem[\protect\citeauthoryear{{Lah{\'e}n}, {Naab}, {Johansson}, {Elmegreen},
  {Hu}  \& {Walch}}{{Lah{\'e}n} et~al.}{2019}]{Lahen2019A}
{Lah{\'e}n} N.,  {Naab} T.,  {Johansson} P.~H.,  {Elmegreen} B.,  {Hu} C.-Y.,
  {Walch} S.,  2019, \mn@doi [\apjl] {10.3847/2041-8213/ab2a13}, \href
  {https://ui.adsabs.harvard.edu/abs/2019ApJ...879L..18L} {879, L18}

\bibitem[\protect\citeauthoryear{{Lah{\'e}n}, {Naab}, {Johansson}, {Elmegreen},
  {Hu}, {Walch}, {Steinwand el}  \& {Moster}}{{Lah{\'e}n}
  et~al.}{2020}]{Lahen2019B}
{Lah{\'e}n} N.,  {Naab} T.,  {Johansson} P.~H.,  {Elmegreen} B.,  {Hu} C.-Y.,
  {Walch} S.,  {Steinwand el} U.~P.,   {Moster} B.~P.,  2020, \mn@doi [\apj]
  {10.3847/1538-4357/ab7190}, \href
  {https://ui.adsabs.harvard.edu/abs/2020ApJ...891....2L} {891, 2}

\bibitem[\protect\citeauthoryear{{Lee}, {Joo}, {Sohn}, {Rey}, {Lee}  \&
  {Walker}}{{Lee} et~al.}{1999}]{Lee1999}
{Lee} Y.~W.,  {Joo} J.~M.,  {Sohn} Y.~J.,  {Rey} S.~C.,  {Lee} H.~C.,
  {Walker} A.~R.,  1999, \mn@doi [\nat] {10.1038/46985}, \href
  {https://ui.adsabs.harvard.edu/abs/1999Natur.402...55L} {402, 55}

\bibitem[\protect\citeauthoryear{{Leitherer} et~al.,}{{Leitherer}
  et~al.}{1999}]{Leitherer1999}
{Leitherer} C.,  et~al., 1999, \mn@doi [\apjs] {10.1086/313233}, \href
  {http://adsabs.harvard.edu/abs/1999ApJS..123....3L} {123, 3}

\bibitem[\protect\citeauthoryear{{Longmore} et~al.,}{{Longmore}
  et~al.}{2014}]{Longmore2014}
{Longmore} S.~N.,  et~al., 2014, \mn@doi [Protostars and Planets VI]
  {10.2458/azu_uapress_9780816531240-ch013}, \href
  {http://adsabs.harvard.edu/abs/2014prpl.conf..291L} {pp 291--314}

\bibitem[\protect\citeauthoryear{{Marino}, {Villanova}, {Piotto}, {Milone},
  {Momany}, {Bedin}  \& {Medling}}{{Marino} et~al.}{2008}]{Marino2008}
{Marino} A.~F.,  {Villanova} S.,  {Piotto} G.,  {Milone} A.~P.,  {Momany} Y.,
  {Bedin} L.~R.,   {Medling} A.~M.,  2008, \mn@doi [\aap]
  {10.1051/0004-6361:200810389}, \href
  {http://adsabs.harvard.edu/abs/2008A%26A...490..625M} {490, 625}

\bibitem[\protect\citeauthoryear{{Mart{\'{\i}}nez-Gonz{\'a}lez}, {Silich}  \&
  {Tenorio-Tagle}}{{Mart{\'{\i}}nez-Gonz{\'a}lez} et~al.}{2014}]{SergioMG2014}
{Mart{\'{\i}}nez-Gonz{\'a}lez} S.,  {Silich} S.,   {Tenorio-Tagle} G.,  2014,
  \mn@doi [\apj] {10.1088/0004-637X/785/2/164}, \href
  {http://adsabs.harvard.edu/abs/2014ApJ...785..164M} {785, 164}

\bibitem[\protect\citeauthoryear{{Mart{\'{\i}}nez-Gonz{\'a}lez},
  {Tenorio-Tagle}  \& {Silich}}{{Mart{\'{\i}}nez-Gonz{\'a}lez}
  et~al.}{2016}]{Sergio2016}
{Mart{\'{\i}}nez-Gonz{\'a}lez} S.,  {Tenorio-Tagle} G.,   {Silich} S.,  2016,
  \mn@doi [\apj] {10.3847/0004-637X/816/1/39}, \href
  {https://ui.adsabs.harvard.edu/abs/2016ApJ...816...39M} {816, 39}

\bibitem[\protect\citeauthoryear{{Mart{\'{\i}}nez-Gonz{\'a}lez}, {W{\"u}nsch}
  \& {Palou{\v s}}}{{Mart{\'{\i}}nez-Gonz{\'a}lez} et~al.}{2017}]{Sergio2017}
{Mart{\'{\i}}nez-Gonz{\'a}lez} S.,  {W{\"u}nsch} R.,   {Palou{\v s}} J.,  2017,
  \mn@doi [\apj] {10.3847/1538-4357/aa7510}, \href
  {https://ui.adsabs.harvard.edu/abs/2017ApJ...843...95M} {843, 95}

\bibitem[\protect\citeauthoryear{{Mirabel}}{{Mirabel}}{2017a}]{Mirabel2017A}
{Mirabel} F.,  2017a, \mn@doi [\nar] {10.1016/j.newar.2017.04.002}, \href
  {http://adsabs.harvard.edu/abs/2017NewAR..78....1M} {78, 1}

\bibitem[\protect\citeauthoryear{{Mirabel}}{{Mirabel}}{2017b}]{Mirabel2017B}
{Mirabel} I.~F.,  2017b, in {Gomboc} A.,  ed.,  IAU Symposium Vol. 324, New
  Frontiers in Black Hole Astrophysics. pp 303--306 (\mn@eprint {arXiv}
  {1611.09266}), \mn@doi{10.1017/S1743921316012904}

\bibitem[\protect\citeauthoryear{{Monreal-Ibero}, {V{\'\i}lchez}, {Walsh}  \&
  {Mu{\~n}oz-Tu{\~n}{\'o}n}}{{Monreal-Ibero} et~al.}{2010}]{MonrealIbero2010}
{Monreal-Ibero} A.,  {V{\'\i}lchez} J.~M.,  {Walsh} J.~R.,
  {Mu{\~n}oz-Tu{\~n}{\'o}n} C.,  2010, \mn@doi [\aap]
  {10.1051/0004-6361/201014154}, \href
  {https://ui.adsabs.harvard.edu/abs/2010A&A...517A..27M} {517, A27}

\bibitem[\protect\citeauthoryear{{Nisini}, {Antoniucci}, {Alcal{\'a}},
  {Giannini}, {Manara}, {Natta}, {Fedele}  \& {Biazzo}}{{Nisini}
  et~al.}{2018}]{Nisini2018}
{Nisini} B.,  {Antoniucci} S.,  {Alcal{\'a}} J.~M.,  {Giannini} T.,  {Manara}
  C.~F.,  {Natta} A.,  {Fedele} D.,   {Biazzo} K.,  2018, \mn@doi [\aap]
  {10.1051/0004-6361/201730834}, \href
  {https://ui.adsabs.harvard.edu/abs/2018A%26A...609A..87N} {609, A87}

\bibitem[\protect\citeauthoryear{{Oey}, {Herrera}, {Silich}, {Reiter}, {James},
  {Jaskot}  \& {Micheva}}{{Oey} et~al.}{2017}]{Oey2017}
{Oey} M.~S.,  {Herrera} C.~N.,  {Silich} S.,  {Reiter} M.,  {James} B.~L.,
  {Jaskot} A.~E.,   {Micheva} G.,  2017, \mn@doi [\apjl]
  {10.3847/2041-8213/aa9215}, \href
  {http://adsabs.harvard.edu/abs/2017ApJ...849L...1O} {849, L1}

\bibitem[\protect\citeauthoryear{{Owen}, {Ercolano}  \& {Clarke}}{{Owen}
  et~al.}{2011}]{Owen2011}
{Owen} J.~E.,  {Ercolano} B.,   {Clarke} C.~J.,  2011, \mn@doi [\mnras]
  {10.1111/j.1365-2966.2010.17818.x}, \href
  {https://ui.adsabs.harvard.edu/abs/2011MNRAS.412...13O} {412, 13}

\bibitem[\protect\citeauthoryear{{Portegies Zwart}, {McMillan}  \&
  {Gieles}}{{Portegies Zwart} et~al.}{2010}]{Zwart2010}
{Portegies Zwart} S.~F.,  {McMillan} S.~L.~W.,   {Gieles} M.,  2010, \mn@doi
  [\araa] {10.1146/annurev-astro-081309-130834}, \href
  {http://adsabs.harvard.edu/abs/2010ARA%26A..48..431P} {48, 431}

\bibitem[\protect\citeauthoryear{{Rahner}, {Pellegrini}, {Glover}  \&
  {Klessen}}{{Rahner} et~al.}{2017}]{Rahner2017}
{Rahner} D.,  {Pellegrini} E.~W.,  {Glover} S. C.~O.,   {Klessen} R.~S.,  2017,
  \mn@doi [\mnras] {10.1093/mnras/stx1532}, \href
  {https://ui.adsabs.harvard.edu/abs/2017MNRAS.470.4453R} {470, 4453}

\bibitem[\protect\citeauthoryear{{Raymond}, {Cox}  \& {Smith}}{{Raymond}
  et~al.}{1976}]{Raymond1976}
{Raymond} J.~C.,  {Cox} D.~P.,   {Smith} B.~W.,  1976, \mn@doi [\apj]
  {10.1086/154170}, \href {http://adsabs.harvard.edu/abs/1976ApJ...204..290R}
  {204, 290}

\bibitem[\protect\citeauthoryear{{Renzini} et~al.,}{{Renzini}
  et~al.}{2015}]{Renzini2015}
{Renzini} A.,  et~al., 2015, \mn@doi [\mnras] {10.1093/mnras/stv2268}, \href
  {http://adsabs.harvard.edu/abs/2015MNRAS.454.4197R} {454, 4197}

\bibitem[\protect\citeauthoryear{{Richling} \& {Yorke}}{{Richling} \&
  {Yorke}}{2000}]{Richling2000}
{Richling} S.,  {Yorke} H.~W.,  2000, \mn@doi [\apj] {10.1086/309198}, \href
  {https://ui.adsabs.harvard.edu/abs/2000ApJ...539..258R} {539, 258}

\bibitem[\protect\citeauthoryear{{Seale}, {Looney}, {Wong}, {Ott}, {Klein}  \&
  {Pineda}}{{Seale} et~al.}{2012}]{Seale2012}
{Seale} J.~P.,  {Looney} L.~W.,  {Wong} T.,  {Ott} J.,  {Klein} U.,   {Pineda}
  J.~L.,  2012, \mn@doi [\apj] {10.1088/0004-637X/751/1/42}, \href
  {https://ui.adsabs.harvard.edu/abs/2012ApJ...751...42S} {751, 42}

\bibitem[\protect\citeauthoryear{{Silich} \& {Tenorio-Tagle}}{{Silich} \&
  {Tenorio-Tagle}}{2017}]{Silich2017}
{Silich} S.,  {Tenorio-Tagle} G.,  2017, \mn@doi [\mnras]
  {10.1093/mnras/stw2879}, \href
  {http://adsabs.harvard.edu/abs/2017MNRAS.465.1375S} {465, 1375}

\bibitem[\protect\citeauthoryear{{Silich} \& {Tenorio-Tagle}}{{Silich} \&
  {Tenorio-Tagle}}{2018}]{Silich2018}
{Silich} S.,  {Tenorio-Tagle} G.,  2018, \mn@doi [\mnras]
  {10.1093/mnras/sty1383}, \href
  {http://adsabs.harvard.edu/abs/2018MNRAS.478.5112S} {478, 5112}

\bibitem[\protect\citeauthoryear{{Silich}, {Tenorio-Tagle}  \&
  {Rodr{\'{\i}}guez-Gonz{\'a}lez}}{{Silich} et~al.}{2004}]{Silich2004}
{Silich} S.,  {Tenorio-Tagle} G.,   {Rodr{\'{\i}}guez-Gonz{\'a}lez} A.,  2004,
  \mn@doi [\apj] {10.1086/421702}, \href
  {http://adsabs.harvard.edu/abs/2004ApJ...610..226S} {610, 226}

\bibitem[\protect\citeauthoryear{{Smith} \& {Gallagher}}{{Smith} \&
  {Gallagher}}{2001}]{Smith2001}
{Smith} L.~J.,  {Gallagher} J.~S.,  2001, \mn@doi [\mnras]
  {10.1046/j.1365-8711.2001.04627.x}, \href
  {http://adsabs.harvard.edu/abs/2001MNRAS.326.1027S} {326, 1027}

\bibitem[\protect\citeauthoryear{{Smith}, {Crowther}, {Calzetti}  \&
  {Sidoli}}{{Smith} et~al.}{2016}]{Smith2016}
{Smith} L.~J.,  {Crowther} P.~A.,  {Calzetti} D.,   {Sidoli} F.,  2016, \mn@doi
  [\apj] {10.3847/0004-637X/823/1/38}, \href
  {http://adsabs.harvard.edu/abs/2016ApJ...823...38S} {823, 38}

\bibitem[\protect\citeauthoryear{{St{\"o}rzer} \& {Hollenbach}}{{St{\"o}rzer}
  \& {Hollenbach}}{1999}]{Storzer1999}
{St{\"o}rzer} H.,  {Hollenbach} D.,  1999, \mn@doi [\apj] {10.1086/307055},
  \href {https://ui.adsabs.harvard.edu/abs/1999ApJ...515..669S} {515, 669}

\bibitem[\protect\citeauthoryear{{Sz{\'e}csi} \& {W{\"u}nsch}}{{Sz{\'e}csi} \&
  {W{\"u}nsch}}{2019}]{Szecsi2019}
{Sz{\'e}csi} D.,  {W{\"u}nsch} R.,  2019, \mn@doi [\apj]
  {10.3847/1538-4357/aaf4be}, \href
  {https://ui.adsabs.harvard.edu/abs/2019ApJ...871...20S} {871, 20}

\bibitem[\protect\citeauthoryear{{Tenorio-Tagle}, {Mu{\~n}oz-Tu{\~n}{\'o}n},
  {Silich}  \& {Cassisi}}{{Tenorio-Tagle} et~al.}{2015}]{TenorioTagle2015}
{Tenorio-Tagle} G.,  {Mu{\~n}oz-Tu{\~n}{\'o}n} C.,  {Silich} S.,   {Cassisi}
  S.,  2015, \mn@doi [\apjl] {10.1088/2041-8205/814/1/L8}, \href
  {http://adsabs.harvard.edu/abs/2015ApJ...814L...8T} {814, L8}

\bibitem[\protect\citeauthoryear{{Tenorio-Tagle}, {Silich}, {Palous},
  {Mu{\~n}oz-Tu{\~n}{\'o}n}  \& {Wunsch}}{{Tenorio-Tagle}
  et~al.}{2019}]{TenorioTagle2019}
{Tenorio-Tagle} G.,  {Silich} S.,  {Palous} J.,  {Mu{\~n}oz-Tu{\~n}{\'o}n} C.,
   {Wunsch} R.,  2019, arXiv e-prints, \href
  {https://ui.adsabs.harvard.edu/abs/2019arXiv190510183T} {}

\bibitem[\protect\citeauthoryear{{Thuan}, {Sauvage}  \& {Madden}}{{Thuan}
  et~al.}{1999}]{Thuan1999}
{Thuan} T.~X.,  {Sauvage} M.,   {Madden} S.,  1999, \mn@doi [\apj]
  {10.1086/307152}, \href
  {https://ui.adsabs.harvard.edu/abs/1999ApJ...516..783T} {516, 783}

\bibitem[\protect\citeauthoryear{{Turner} \& {Beck}}{{Turner} \&
  {Beck}}{2004}]{Turner2004}
{Turner} J.~L.,  {Beck} S.~C.,  2004, \mn@doi [\apjl] {10.1086/382699}, \href
  {https://ui.adsabs.harvard.edu/abs/2004ApJ...602L..85T} {602, L85}

\bibitem[\protect\citeauthoryear{{Turner}, {Beck}  \& {Ho}}{{Turner}
  et~al.}{2000}]{Turner2000}
{Turner} J.~L.,  {Beck} S.~C.,   {Ho} P.~T.~P.,  2000, \mn@doi [\apjl]
  {10.1086/312586}, \href {http://adsabs.harvard.edu/abs/2000ApJ...532L.109T}
  {532, L109}

\bibitem[\protect\citeauthoryear{{Turner}, {Beck}, {Benford}, {Consiglio},
  {Ho}, {Kov{\'a}cs}, {Meier}  \& {Zhao}}{{Turner} et~al.}{2015}]{Turner2015}
{Turner} J.~L.,  {Beck} S.~C.,  {Benford} D.~J.,  {Consiglio} S.~M.,  {Ho}
  P.~T.~P.,  {Kov{\'a}cs} A.,  {Meier} D.~S.,   {Zhao} J.-H.,  2015, \mn@doi
  [\nat] {10.1038/nature14218}, \href
  {http://adsabs.harvard.edu/abs/2015Natur.519..331T} {519, 331}

\bibitem[\protect\citeauthoryear{{Turner}, {Consiglio}, {Beck}, {Goss}, {Ho},
  {Meier}, {Silich}  \& {Zhao}}{{Turner} et~al.}{2017}]{Turner2017}
{Turner} J.~L.,  {Consiglio} S.~M.,  {Beck} S.~C.,  {Goss} W.~M.,  {Ho}
  P.~T.~P.,  {Meier} D.~S.,  {Silich} S.,   {Zhao} J.-H.,  2017, \mn@doi [\apj]
  {10.3847/1538-4357/aa8669}, \href
  {http://adsabs.harvard.edu/abs/2017ApJ...846...73T} {846, 73}

\bibitem[\protect\citeauthoryear{{Vacca}, {Johnson}  \& {Conti}}{{Vacca}
  et~al.}{2002}]{Vacca2002}
{Vacca} W.~D.,  {Johnson} K.~E.,   {Conti} P.~S.,  2002, \mn@doi [\aj]
  {10.1086/338644}, \href
  {https://ui.adsabs.harvard.edu/abs/2002AJ....123..772V} {123, 772}

\bibitem[\protect\citeauthoryear{{Vanzi} \& {Sauvage}}{{Vanzi} \&
  {Sauvage}}{2004}]{Vanzi2004}
{Vanzi} L.,  {Sauvage} M.,  2004, \mn@doi [\aap] {10.1051/0004-6361:20034635},
  \href {https://ui.adsabs.harvard.edu/abs/2004A%26A...415..509V} {415, 509}

\bibitem[\protect\citeauthoryear{{Walker}, {Longmore}, {Bastian}, {Kruijssen},
  {Rathborne}, {Galv{\'a}n-Madrid}  \& {Liu}}{{Walker}
  et~al.}{2016}]{Walker2016}
{Walker} D.~L.,  {Longmore} S.~N.,  {Bastian} N.,  {Kruijssen} J.~M.~D.,
  {Rathborne} J.~M.,  {Galv{\'a}n-Madrid} R.,   {Liu} H.~B.,  2016, \mn@doi
  [\mnras] {10.1093/mnras/stw313}, \href
  {https://ui.adsabs.harvard.edu/abs/2016MNRAS.457.4536W} {457, 4536}

\bibitem[\protect\citeauthoryear{{Whitmore} et~al.,}{{Whitmore}
  et~al.}{2011}]{Whitmore2011}
{Whitmore} B.~C.,  et~al., 2011, \mn@doi [\apj] {10.1088/0004-637X/729/2/78},
  \href {https://ui.adsabs.harvard.edu/abs/2011ApJ...729...78W} {729, 78}

\bibitem[\protect\citeauthoryear{{W{\"u}nsch}, {Palou{\v s}}, {Tenorio-Tagle}
  \& {Ehlerov{\'a}}}{{W{\"u}nsch} et~al.}{2017}]{Wunsch2017}
{W{\"u}nsch} R.,  {Palou{\v s}} J.,  {Tenorio-Tagle} G.,   {Ehlerov{\'a}} S.,
  2017, \mn@doi [\apj] {10.3847/1538-4357/835/1/60}, \href
  {http://adsabs.harvard.edu/abs/2017ApJ...835...60W} {835, 60}

\makeatother
\end{thebibliography}

\bsp	
\label{lastpage}
\end{document}